\documentclass[prd,superscriptaddress,showpacs,nofootinbib,amsmath,amssymb]{revtex4}

\usepackage{bm}
\usepackage{amsfonts}
\usepackage{latexsym}
\usepackage[latin1]{inputenc}
\usepackage{graphicx}
\usepackage{amsmath}

\usepackage{booktabs}
\usepackage{dcolumn}

\def\jnl@style{\it}
\def\aaref@jnl#1{{\jnl@style#1}}

\def\aaref@jnl#1{{\jnl@style#1}}

\def\aj{\aaref@jnl{AJ}}                   
\def\apj{\aaref@jnl{ApJ}}                 
\def\apjl{\aaref@jnl{ApJ}}                
\def\apjs{\aaref@jnl{ApJS}}               
\def\apss{\aaref@jnl{Ap\&SS}}             
\def\aap{\aaref@jnl{A\&A}}                
\def\aapr{\aaref@jnl{A\&A~Rev.}}          
\def\aaps{\aaref@jnl{A\&AS}}              
\def\mnras{\aaref@jnl{MNRAS}}             
\def\prd{\aaref@jnl{Phys.~Rev.~D}}        
\def\prl{\aaref@jnl{Phys.~Rev.~Lett.}}    
\def\qjras{\aaref@jnl{QJRAS}}             
\def\skytel{\aaref@jnl{S\&T}}             
\def\ssr{\aaref@jnl{Space~Sci.~Rev.}}     
\def\zap{\aaref@jnl{ZAp}}                 
\def\nat{\aaref@jnl{Nature}}              
\def\aplett{\aaref@jnl{Astrophys.~Lett.}} 
\def\apspr{\aaref@jnl{Astrophys.~Space~Phys.~Res.}} 
\def\physrep{\aaref@jnl{Phys.~Rep.}}      
\def\physscr{\aaref@jnl{Phys.~Scr}}       
\def\commat{\aaref@jnl{Comm.~Math.~Phys.}}		
\def\science{\aaref@jnl{Science}}		
\def\cqg{\aaref@jnl{Class.~Quantum Gravity}}		
\def\jpcs{\aaref@jnl{JPCS}}					
\def\ijmp{\aaref@jnl{Int.~J.~Mod.~Phys.}}			

\newcommand{\unit}[1]{\ensuremath{\, \mathrm{#1}}}
\allowdisplaybreaks[1]

\begin{document}

\title{Oscillations and instabilities of fast and differentially rotating relativistic stars}

\author{Christian Kr\"uger}
\author{Erich Gaertig}
\affiliation{Theoretical Astrophysics, Eberhard-Karls University of T\"ubingen, T\"ubingen 72076, Germany}
\author{Kostas D. Kokkotas}
\affiliation{Theoretical Astrophysics, Eberhard-Karls University of T\"ubingen, T\"ubingen 72076, Germany}
\affiliation{Department of Physics, Aristotle University of Thessaloniki, Thessaloniki 54124, Greece}
\date{\today}

\begin{abstract}
We study non-axisymmetric oscillations of rapidly and differentially rotating
relativistic stars in the Cowling approximation. Our equilibrium models are
sequences of relativistic polytropes, where the differential rotation is
described by the relativistic $j$-constant law. We show that a small degree of
differential rotation raises the critical rotation value for which the
quadrupolar f-mode becomes prone to the CFS instability, while the critical
value of $T/|W|$ at the mass-shedding limit is raised even more. For
stiffer equations of state these effects are even more pronounced. When
increasing differential rotation further to a high degree, the neutral point
of the CFS instability first reaches a local maximum and is lowered
afterwards. For stars with a rather high compactness we find that for a large
degree of differential rotation the absolute value of the critical $T/|W|$ is
below the corresponding value for rigid rotation. We conclude that the onset
of the CFS instability is eased for
a small degree of differential rotation and for a large degree at
least in stars with a higher compactness.
Moreover, we were able to extract the eigenfrequencies and the eigenfunctions
of r-modes for differentially rotating stars and our simulations show a good
qualitative agreement with previous Newtonian results. 
\end{abstract}


\maketitle

\section{Introduction}
\label{sec:introduction}

Rapidly spinning neutron stars can be destabilized due to the CFS 
\cite{1970ApJ...161..561C,1978ApJ...221..937F,1978ApJ...222..281F}
instability. This instability, for quadrupole non-axisymmetric perturbations
with $m=2$, sets in for uniformly rotating polytropes when $\beta=T/|W|\approx
0.14$ where $T$ is the rotational kinetic energy and $W$ is the gravitational
potential energy of the star. For higher values of $m$ the instability sets in
for smaller rotational rates but the growth time is considerable longer and
viscosity stabilizes the star. The CFS instability is generic for
r-modes, i.\,e. these modes are CFS unstable for any rotational rate while
for the f-modes the previous value of $\beta$ is approximately correct.
This value of $\beta$ corresponds to rotational frequencies of the stars which
are 85-90\% of the Kepler frequency. These high spin frequencies have not yet been observed in nature but it is expected that they may be met in newly born neutron stars.  Newly born neutron stars are expected to rotate not only fast but also differentially, and this is a factor that has to be taken into account in the study of instabilities.

Actually, differential rotation can appear in many phases of the stellar evolution of a neutron star, such as in protoneutron
stars~\cite{2002A&A...393..523D,2004ApJ...600..834O}, in the massive remnant of binary neutron star
mergers~\cite{1994ApJ...432..242R,Shibata:1999wm,Rasio:1999zr}, or as a results of
stellar oscillations (r-modes) that may drive the star into
differential rotation via nonliner
effects~\cite{
2000ApJ...531L.139R,
2001MNRAS.324..917L,
2001PhRvL..86.1148S,
2004PhRvD..69h4001S}.
The phase where the star rotates differentially  can last between seconds to months, depending on the dissipative mechanism that drive the star to uniform
rotation, such as viscosity, magnetic braking~\cite{2000ApJ...544..397S,2003ApJ...599.1272C} and turbulent motion~\cite{1977ApJ...217..244H,liu:044009}.
However, this very short period is met during the most violent phases of neutron star's life, as in
the case of a core collapse or binary merging. This is exactly the
time when we expect to get the strongest emission in gravitational
waves.
 Since the ground based
detectors are reaching sensitivities which allow the detection of
gravitational wave signals from oscillating or unstable neutron stars
the critical point for the onset of instabilities, the growth times of the instabilities and the exact frequencies of the emitted waves are urgently needed.

In this work we study within the general relativistic framework the oscillations
and instabilities of fast and differentially rotating neutron stars in the so
called Cowling approximation i.e. assuming a fixed spacetime. This is the
first study of its kind since earlier works were using certain approximations
e.g  Newtonian theory \cite{2002ApJ...578..413K} or slow rotation \cite{2007PhRvD..75f4019S,2008PhRvD..77b4029P}. Results, from fully nonlinear calculations exist  only for the axisymmetric oscillation ~\cite{Stergioulas:2003ep,Dimmelmeier:2005zk}.  Finally, the f-mode and the secular stability limits have been investigated
in~\cite{2002ApJ...568L..41Y}.

Recently, it has been found that in stars that rotate with a high degree of differential rotation an $m=2$ dynamical instability can
appear even for considerably low rotation rates ($T/|W|\sim
O(10^{-2})$) as suggested
in~\cite{2002MNRAS.334L..27S,2005PhRvD..71b4014S}.   In addition, an $m=1$ 
dynamical instability has been identified for high degrees of
differential rotation and soft equations of state
\cite{2001ApJ...550L.193C,2003ApJ...595..352S}.  More recently, the discovery of $m=1$ and $m=2$ dynamical instability
even for stiff equations of state \cite{2006ApJ...651.1068O} has been reported .
Studies based on linear analysis
\cite{2005ApJ...618L..37W,2006MNRAS.368.1429S} suggest that low
$T/|W|$ instabilities might be triggered when the corotation points of
the unstable modes fall within the differentially rotating structure
of the star.

The structure of the paper is as follows. In Section~\ref{sec:formulation} 
we present the formulation that we have used to construct the equilibrium configurations.
We then derive the perturbation equations
in the general case of barotropic and non-axisymmetric
perturbations, which are numerically solved in
Section~\ref{sec:results} for the non-axisymmetric and barotropic case.
Finally, in Section \ref{sec:summary}, we summarize the crucial results of this study.

Throughout the paper we use geometrical units $c=G=1$.

\section{Formulation of the Problem}
\label{sec:formulation}

In this section we first show the basic principles of our problem. We then
derive the evolution equations governing oscillations of perturbations on
compact objects and the appropriate boundary conditions. Next, we present the
equations of state we use for constructing equilibrium models and what kind of 
sequences we will consider. Finally, we explain how we solve the problem
numerically.

\subsection{Basic Principles}
\label{ssec:basics}

We study linear perturbations of rapidly and differentially rotating, relativistic neutron stars on a fixed background. First of all, this
requires the computation of equilibrium configurations which are prescribed by their general-relativistic line-element and the specific form of the energy-momentum tensor. This last quantity describes the proper characteristics of the neutron star fluid and therefore, a certain equation of state has to be specified as well; we will address this issue later in this section. 

The background metric of a fast rotating, relativistic star has to be stationary and axisymmetric and using spherical coordinates $(r,\theta,\phi)$, the line-element can be written as
\begin{equation}
    ds^2 = - e^{2\nu}dt^2
           + e^{2\psi} r^2 \sin^2\theta (d\phi - \omega dt)^2
           + e^{2\mu}(dr^2 + r^2 d\theta^2)\,.
\end{equation}
The four unknown functions $\psi$, $\mu$, $\nu$ and $\omega$ are the metric potentials
which depend on $r$ and $\theta$ only and can be determined by solving the time-independent Einstein equations. Furthermore, for equilibrium models the radial and the
polar component of the fluid's four-velocity $u^{\mu}$ vanishes while the remaining two
components are related by the angular velocity $\Omega$ via
\begin{equation}
    \left(u^t,u^r,u^{\theta},u^{\phi}\right)
        = \left(u^t,0,0,\Omega u^t\right)\,.
\end{equation}

In general, the angular velocity $\Omega$ is a function of $r$ and $\theta$ and although the exact form of
$\Omega(r,\theta)$ is arbitrary to some extent, we adopt the so-called
relativistic $j$-constant rotation law
\begin{equation}
\label{eq:diffRotLaw}
    A^2\left(\Omega_c - \Omega\right)
        = \frac{     (\Omega-\omega)   r^2 \sin^2\theta e^{2(\psi-\nu)} }
               { 1 - (\Omega-\omega)^2 r^2 \sin^2\theta e^{2(\psi-\nu)} }\,,
\end{equation}
which satisfies the Rayleigh criterion for local dynamical stability against axisymmetric perturbations and which is commonly used in similar studies.  \cite{Komatsu1989MNRASA,Komatsu1989MNRASB, Stergioulas:2003ep}.
Here, $\Omega_c$ is the angular velocity on the rotation axis and the free parameter $A$ controls
the degree of differential rotation. This quantity has units of length in the SI-system and specifies the
length scale over which the angular velocity varies inside the star. As one can see from equation \eqref{eq:diffRotLaw}, the star is rotating uniformly again when $A$ tends to infinity.

This definition of $A$ is not convenient when considering sequences of equilibrium models with increasing rotation rate, since the radius of the star can vary by a factor of two
which would be reflected in different degrees of differential rotation for each
model. This issue can be circumvented by using a slightly modified parameter
\begin{equation}
    \hat{A}:=A/r_e\,,
\end{equation}
which normalizes $A$ by the equatorial radius $r_e$ of the star; see also \cite{Stergioulas:2003ep}.
This definition maintains the same degree of differential rotation for all
models along a sequence.

We furthermore assume the neutron star to be a perfect fluid without viscosity. In this case, the energy-momentum tensor has the form
\begin{equation}
    T^{\mu\nu} = (\epsilon + p)u^{\mu}u^{\nu} +p g^{\mu\nu}\,,
\end{equation}
where $\epsilon$ is the energy density, $p$ is the pressure and $g^{\mu\nu}$ is the inverse metric. Energy and pressure are not independent quantities but they are related by an equation of state. In this study, we will focus on polytropic equations of state in the form of
\begin{align}
    p        & = K\rho^{\Gamma}\,,\\
    \epsilon & = \rho + Np\,.\nonumber
\end{align}
Here, $\rho$ is the rest-mass density, $K$ is the polytropic constant, $N$ is
the polytropic index and $\Gamma=1 + 1/N$. For barotropic oscillations, pressure- and density-perturbations are connected by the speed of sound; $\delta p = c_s^2\delta\epsilon$. In the case of polytropic equations of state, its value can be computed analytically and it is
\begin{equation}
    c^2_s  = \frac{\Gamma p}{\epsilon + p}.
\end{equation}

The thorough study of neutron star oscillations is very demanding and requires
to solve the full, non-linear equations of general relativity. At this point,
we will introduce some approximations in order to deal with a smaller set of
equations and unknowns. First of all, we restrict the study to small
perturbations around the equilibrium which allows us to linearize the
equations. Second, we will use the so-called relativistic Cowling
approximation, in which all metric perturbations are neglected. This
effectively discards the space-time degrees of freedom and one only has to
deal with the equations governing the motion of the fluid, which follow from
the linear variation of the local law of energy-momentum-conservation
\begin{equation}
\label{eq:conservationLaw}
    \delta\left( \nabla_{\nu}T^{\mu\nu} \right) = 0.
\end{equation}
Since in the Cowling approximation, the covariant derivative is not affected by the fluid perturbation, equation \eqref{eq:conservationLaw} can be written as
\begin{equation}
     \partial_{\nu} \delta T^{\mu\nu}
          + \Gamma^{\mu}_{\gamma\nu} \delta T^{\gamma\nu}
          + \Gamma^{\nu}_{\gamma\nu} \delta T^{\mu\gamma}
        = 0\,,
    \label{equ:nabla_t_eq_0}
\end{equation}
where $\delta T^{\mu\nu}$ is the perturbed energy-momentum tensor.

\subsection{Perturbations Equations}
\label{ssec:perturbationEquations}
The perturbation equations can be derived from equation \eqref{equ:nabla_t_eq_0} and for this, we will follow a very convenient  formulation developed in \cite{2005JPhCS...8...71K,Vavoul2007}. Instead of using the canonical fluid perturbation variables like velocity- or pressure-perturbation, the independent components of the perturbed energy-momentum tensor are utilized for the time-evolution.

To be more specific, we are directly evolving the variables $Q_1:=\delta T^{tt}$, $Q_2:=\delta T^{t\phi}$,
$Q_3:=\delta T^{tr}$ and $Q_4:=\delta T^{t\theta}$ in time, using equation \eqref{equ:nabla_t_eq_0} to obtain the corresponding evolution equations. Additionally, we make use of an auxiliary variable $Q_6:=\delta T^{rr}$ which turns out to be a linear combination of $Q_1$ and $Q_6$ and which helps us in writing the resulting set of differential equations in a more compact way.

The final set of evolution equations is then given by 
\begin{subequations}
\begin{align}
    \begin{split}
    \partial_t Q_1 =
      & - \partial_{\phi}Q_2 - \partial_r Q_3 - \partial_{\theta} Q_4 \\
      & - \left[ e^{-2\nu+2\psi} r^2 \sin^2 \theta (\Omega - \omega)
                         \partial_r\omega
                 + \partial_r\psi + 2\partial_r\mu
                 + 3\partial_r\nu + \frac{2}{r}
          \right] Q_3 \\
      & - \left[ e^{-2\nu+2\psi} r^2 \sin^2 \theta (\Omega - \omega)
                         \partial_{\theta}\omega
                 + \partial_{\theta}\psi + 2\partial_{\theta}\mu
                 + 3\partial_{\theta}\nu + \frac{\cos\theta}{\sin\theta}
          \right] Q_4 \\
      \label{equ:perturb_eq_q1}
    \end{split} \\
    \begin{split}
    \partial_t Q_2 ={}
      & \Omega^2 \partial_{\phi} Q_1
         - 2\Omega\partial_{\phi} Q_2
         - \Omega\partial_r Q_3
         - \Omega\partial_{\theta} Q_4
        + \left[ e^{-2\nu+2\mu}(\Omega-\omega)^2
                 - \frac{e^{-2\psi+2\mu}}{r^2\sin^2\theta}
          \right] \partial_{\phi} Q_6 \\
      & - \left[ \vphantom{\frac{2}{r}} 
                 e^{-2\nu+2\psi} r^2 \sin^2\theta (\Omega-\omega)
                   \omega\partial_r\omega
                 + \partial_r\left(\Omega - \omega\right)
                 + (3\Omega - 2\omega)\partial_r\psi
          \right. \\
      & \qquad
          \left. + 2\Omega\partial_r\mu
                 + (\Omega + 2\omega)\partial_r\nu
                 + (2\Omega - \omega)\frac{2}{r}
          \right] Q_3 \\
      & - \left[ \vphantom{\frac13} 
                 e^{-2\nu+2\psi} r^2 \sin^2\theta (\Omega-\omega)
                   \omega\partial_{\theta}\omega
                 + \partial_{\theta}\left(\Omega - \omega\right)
                 + (3\Omega - 2\omega)\partial_{\theta}\psi
          \right. \\
      & \qquad
          \left. + 2\Omega\partial_{\theta}\mu
                 + (\Omega + 2\omega)\partial_{\theta}\nu
                 + (3\Omega - 2\omega) \frac{\cos\theta}{\sin\theta}
          \right] Q_4 \\
      \label{equ:perturb_eq_q2}
    \end{split} \\
    \begin{split}
    \partial_t Q_3 =
        & - \Omega\partial_{\phi}Q_3 - \partial_r Q_6 \\
        & - \left[ e^{-2\mu+2\nu}\partial_r\nu
                   + e^{-2\mu+2\psi}r\sin^2\theta
                         \left[ \left(\Omega^2-\omega^2\right)
                                \left(1+r\partial_r\psi\right)
                                - r\omega\partial_r\omega
                         \right]
            \right] Q_1 \\
        & - \left[ e^{-2\mu+2\psi}r\sin^2\theta
                \left[r\partial_r\omega
                      - 2(\Omega-\omega)(1+r\partial_r\psi)
                \right]
            \right] Q_2 \\
        & - \left[ \partial_r\nu + 2\partial_r\mu
                   + e^{-2\nu+2\psi}r\sin^2\theta
                     \left(\Omega-\omega\right)^2(1+r\partial_r\psi)
            \right] Q_6 \\
      \label{equ:perturb_eq_q3}
    \end{split} \\
    \begin{split}
    \partial_t Q_4 =
        & - \Omega\partial_{\phi}Q_4 - \frac{1}{r^2}\partial_{\theta} Q_6 \\
        & - \frac{1}{r^2}
            \left[ e^{-2\mu+2\nu}\partial_{\theta}\nu
                   + e^{-2\mu+2\psi}r^2\sin\theta
                     \left[\left(\Omega^2-\omega^2\right)
                           \left(\cos\theta+\sin\theta\partial_{\theta}\psi
                           \right)
                           - \sin\theta\omega\partial_{\theta}\omega
                     \right]
            \right] Q_1 \\
        & - \frac{1}{r^2}
            \left[ e^{-2\mu+2\psi}r^2\sin\theta
                \left[\sin\theta\partial_{\theta}\omega
                    - 2\left(\Omega-\omega\right)
                      \left(\cos\theta+\sin\theta\partial_{\theta}\psi\right)
                \right]
            \right] Q_2 \\
         & - \frac{1}{r^2}
            \left[ \partial_{\theta}\nu + 2\partial_{\theta}\mu
                + e^{-2\nu+2\psi}r^2\sin\theta
                    \left(\Omega-\omega\right)^2
                    \left(\cos\theta+\sin\theta\partial_{\theta}\psi\right)
            \right] Q_6 \\
      \label{equ:perturb_eq_q4}
    \end{split}
\end{align}
\end{subequations}
Additionally, the value of $Q_6$ in terms of $Q_1$ and $Q_2$ is calculated  by
\begin{align}
    \begin{split}
    Q_6 = & \frac{ e^{-2\mu+2\nu}c_s^2}
               { 1 - e^{-2\nu+2\psi}r^2\sin^2\theta(\Omega-\omega)^2c_s^2 } \\
          & \times \left[
            \left[ 1+e^{-2\nu+2\psi}r^2\sin^2\theta(\Omega^2-\omega^2) \right]
            Q_1
            - 2e^{-2\nu+2\psi}r^2\sin^2\theta(\Omega-\omega) Q_2
          \right]\,.
    \end{split}
\end{align}

Since equilibrium stars are axisymmetric, but not neccessarily spherically symmetric in case of rapid rotation, we will not use an angular decomposition into spherical harmonics but instead only separate the azimuthal part according to  
\begin{equation*}
    Q_i(t,r,\theta,\phi):=\tilde{Q}_i(t,r,\theta)e^{im\phi}
        \quad \text{for}\quad i=1,\ldots,4.
\end{equation*}
Due to this decomposition, the evolution equations \eqref{equ:perturb_eq_q1}-\eqref{equ:perturb_eq_q4} are only slightly modified and we just have to perform the substitution $\partial_{\phi}\to im$. In the following discussion, we will omit the tilde to avoid confusion.

\subsection{Boundary Conditions}
\label{ssec:boundaryConditions}
In order to close the system of equations \eqref{equ:perturb_eq_q1}-\eqref{equ:perturb_eq_q4}, one needs to impose proper boundary conditions. In the study presented here, this boundary consists of the stellar surface, the equatorial plane and the rotation axis. Special care has to be taken of the origin which is mapped onto a boundary line in spherical coordinates.

Let us first consider the stellar surface, where the radial and polar perturbation variables $Q_3$ and $Q_4$ are zero by virtue of their definition
\begin{eqnarray*}
    Q_3 & = & \delta T^{tr} =  \left( \epsilon+p \right) u^t \delta u^r = 0 \\
    Q_4 & =  & \delta T^{t\theta} = \left( \epsilon+p \right) u^t \delta u^{\theta} = 0\,,
\end{eqnarray*}
because pressure and energy density vanish there. $Q_1$ and $Q_2$ however have to remain continuously differentiable there.

On an axisymmetric background, each oscillation mode can be assigned to one of
two possible parity classes which behave differently under space inversion; see
\cite{1999ApJ...521..764L}. Due to the decomposition with respect to the azimuthal angle $\phi$,
the behaviour under reflection with respect to the equatorial plane is also
well-defined and the two parity classes obey different conditions in the
equatorial plane. For one class of modes, the perturbation variables
$Q_1$, $Q_2$ and $Q_3$ are all even, whereas $Q_4$ is odd. The $l=m$
fundamental mode belongs to this class, since the angular part of the pressure perturbation behaves like the scalar spherical harmonic $Y_l^m$, which is even as well.
In contrast, the perturbation variables of modes belonging to the second class
behave exactly the other way around. The $l=m$ r-mode is an
example for a mode of this specific class. We therefore apply different boundary
conditions along the equatorial plane, depending on which class of modes we want to excite.

For deriving the boundary conditions along the rotation axis and at the origin, we use a representation of pressure and velocity perturbations according to
\cite{2005IJMPD..14..543S}. These expressions were derived in the so-called slow-rotation
approximation where one neglects high order effects of rotation, but the relevant coupling between the two different mode classes is already included within this approach and does not change at all in rapidly rotating stars. The boundary conditions we use in this study for discussing spherically symmetric (mostly for code verification purposes) and quadrupolar perturbations are shown in Tables \ref{tab:bc_origin} and \ref{tab:bc_rotaxis}.

\begin{table}[ht!]
    \centering
    \begin{tabular}{cccccc}
    \hline
      $\left|m\right|$
        & $\left.Q_1\right|_{r = 0}$~~~
                 & ~~~$\left.Q_2\right|_{r = 0}$~~~
                          & ~~~$\left.Q_3\right|_{r = 0}$~~~
                               & ~~~$\left.Q_4\right|_{r = 0}$ \\
    \hline
      0 & finite & finite & 0  & 0 \\
      2 &   0    &    0   & 0  & 0 \\
    \hline
    \end{tabular}
    \caption{Boundary conditions for the perturbation variables at the origin.}
    \label{tab:bc_origin}
\end{table}

\begin{table}[ht!]
    \centering
    \begin{tabular}{cccccc}
    \hline
      $\left|m\right|$
         & $\left.Q_1\right|_{\theta = 0}$~~~
                  & ~~~$\left.Q_2\right|_{\theta = 0}$~~~
                           & ~~~$\left.Q_3\right|_{\theta = 0}$~~~
                                    & ~~~$\left.Q_4\right|_{\theta = 0}$ \\
    \hline
       0 & finite & finite & finite & 0 \\
       2 & $\partial_{\theta}=0$
                  & $\partial_{\theta}=0$
                           &    0   & 0 \\
    \hline
    \end{tabular}
    \caption{Boundary conditions for the perturbation variables along the
      rotation axis.}
    \label{tab:bc_rotaxis}
\end{table}

\subsection{Equilibrium Models}
\label{ssec:equilibriumModels}

For constructing the equilibriums models, we use the \texttt{rns}-code
developed by Stergioulas \cite{rns-v1.1,1995ApJ...444..306S}, which was later extended to handle differential rotation. Furthermore, we are using three different equations of state which were already used for the study of uniformly rotating stars in \cite{PhysRevD.78.064063,Gaertig:2009rm} and for which we construct equilibrium sequences with a constant central rest-mass density. All three EoS are polytropic ones; but two of them are fits to tabulated equations of state and together they cover very well the expected range of neutron star massses and radii, see for example \cite{Ozel:2006fk,Ozel:2008lr}.

For all equations of state, we
consider uniformly rotating models as well as their differentially rotating
counterparts for different values of $\hat{A}$. All
sequences start with a non-rotating model for which the axis ratio $r_p/r_e$
of polar to equatorial coordinate radius equals 1. Subsequent models are
computed by decreasing this ratio by a factor of 0.05 until the Kepler limit
is reached.

The EoS B is a widely used  polytropic equation of state with parameters
$\Gamma=2$ and $K=100$ in units of $G=c=M_\odot = 1$, and here we consider the commonly used BU sequence, for
which the dimensionless central rest-mass density is given by
$\rho_c=1.28\times 10^{-3}$. This leads to standard values for the mass and the radius of the non-rotating model; a
detailed list with properties of the BU models and their differentially
rotating couterparts with $\hat{A}=1$ (B sequence) is given in
\cite{Stergioulas:2003ep}. The non-rotating model, which is labelled BU0, has
a gravitational mass of $M=1.4\,M_{\odot}$ and a circumferential radius of
$R_e=14.16\unit{km}$. The fastest rotating member of the uniformly rotating B sequence has an
axis ratio of $r_p/r_e=0.58$, a slightly increased mass of $M=1.695\,M_{\odot}$
and an equatorial radius of $R=19.96\unit{km}$.

When the star is differentially rotating (the following discussion refers to the B
sequence with $\hat{A}=1$), these changes become
even more evident. Due to the fact that differential rotation allows to store a large amount of angular momentum near the rotation axis, the value of $T/|W|$ can reach much higher values.

Furthermore, the star can become more compact near the rotation axis in comparison
to the corresponding uniformly rotating star which results in a very high
mass. The axis ratio for the most rapidly rotating model B12 can be decreased down to a value of 0.4 with a mass of
$M=2.532\,M_{\odot}$ which is an increase of 50 \% when compared to the uniformly rotating model BU9.

As already mentioned, we additionally study two other equations of state which are polytropic fits to the tabulated EoS A and EoS II, see
\cite{PhysRevD.70.084026, 1985ApJ...291..308D, Arnett:1977yq} for  more details. The
oscillation frequencies of neutron star models governed by these EoS have
already been investigated in \cite{PhysRevD.78.064063}.
The fitting parameters for EoS II are $\Gamma=2.34$ and $K = 1186$ and
$\Gamma=2.46$ and $K=1528$ for EoS A again in units where $G = c = M_\odot =1$. For both EoS, we choose models which are very close
to their maximum mass; e.\,g.~see Figure 9 in \cite{PhysRevD.78.064063}.
Apparently, the models for EoS A and EoS II have smaller radii and larger
masses compared to the B sequences. A detailed list of the differentially rotating background models is shown in
Table \ref{tb:equilibriumModels}. 

\begin{table}[ht!]
  \centering
  \begin{tabular}{ccccccc}
    \hline
    Model & ~~~~$r_p/r_e$~~~~ & ~~~~$\Omega_c$~~~~ & ~~~~$\Omega_e$~~~~ & ~~~~$R_e$~~~~ & ~~~~$M$~~~~ 
                                                   & $T/|W|$ \\
          &           & (kHz)      & (kHz)      & (km)  & ($M_{\odot}$) & \\[0.02in]
    \hline
    B0    & 1.00      &    0.000   &    0.000   & 14.16 & 1.400 & 0.000 \\
    B1    & 0.95      &    3.657   &    1.352   & 14.40 & 1.437 & 0.013 \\
    B2    & 0.90      &    5.226   &    1.917   & 14.65 & 1.478 & 0.026 \\
    B3    & 0.85      &    6.473   &    2.354   & 14.93 & 1.525 & 0.040 \\
    B4    & 0.80      &    7.568   &    2.724   & 15.23 & 1.578 & 0.055 \\
    B5    & 0.75      &    8.581   &    3.053   & 15.55 & 1.640 & 0.070 \\
    B6    & 0.70      &    9.555   &    3.352   & 15.90 & 1.713 & 0.087 \\
    B7    & 0.65      &   10.526   &    3.632   & 16.26 & 1.798 & 0.105 \\
    B8    & 0.60      &   11.537   &    3.899   & 16.63 & 1.899 & 0.124 \\
    B9    & 0.55      &   12.651   &    4.166   & 16.99 & 2.020 & 0.144 \\
    B10   & 0.50      &   13.984   &    4.449   & 17.29 & 2.167 & 0.165 \\
    B11   & 0.45      &   15.772   &    4.785   & 17.42 & 2.341 & 0.186 \\
    B12   & 0.40      &   18.508   &    5.245   & 17.20 & 2.532 & 0.207 \\[0.1in]
    
    II0   & 1.00      &    0.000   &    0.000   & 11.71 & 1.908 & 0.000 \\
    II1   & 0.95      &    6.865   &    1.960   & 11.82 & 1.943 & 0.013 \\
    II2   & 0.90      &    9.935   &    2.796   & 11.93 & 1.982 & 0.026 \\
    II3   & 0.85      &   12.481   &    3.454   & 12.04 & 2.025 & 0.041 \\
    II4   & 0.80      &   14.830   &    4.026   & 12.15 & 2.072 & 0.056 \\
    II5   & 0.75      &   17.131   &    4.549   & 12.25 & 2.124 & 0.071 \\
    II6   & 0.70      &   19.490   &    5.043   & 12.34 & 2.181 & 0.087 \\
    II7   & 0.65      &   22.011   &    5.525   & 12.41 & 2.241 & 0.104 \\[0.1in]
    
    A0    & 1.00      &    0.000   &    0.000   &  9.53 & 1.614 & 0.000 \\
    A1    & 0.95      &    8.751   &    2.431   &  9.61 & 1.644 & 0.013 \\
    A2    & 0.90      &   12.681   &    3.468   &  9.70 & 1.676 & 0.027 \\
    A3    & 0.85      &   15.954   &    4.285   &  9.78 & 1.712 & 0.041 \\
    A4    & 0.80      &   18.989   &    4.997   &  9.87 & 1.751 & 0.056 \\
    A5    & 0.75      &   21.977   &    5.647   &  9.94 & 1.794 & 0.071 \\
    A6    & 0.70      &   25.056   &    6.264   & 10.00 & 1.840 & 0.087 \\
    A7    & 0.65      &   28.360   &    6.865   & 10.04 & 1.888 & 0.103 \\

    \hline
  \end{tabular}
 \caption{Stellar properties of the differentially rotating models with $\hat{A}=1$. $\Omega_c$ and $\Omega_e$ are the angular velocities along the rotation axis and at the equator respectively, $R_e$ is the equatorial circumferential radius and $M$ the gravitational mass.}
  \label{tb:equilibriumModels}
\end{table}

\subsection{Numerical Implementation}
\label{sec:numericalTreatment}

After decomposing the perturbed quantities with respect to $\phi$, the
resulting evolution equations form a two-dimensional problem in the spherical
coordinates $r$ and $\theta$. As radial grid coordinate we choose a mapping of the form $s = 0.5 \left(r/r_e\right)^2$, which allows for
stable time evolutions with much less artifical viscosity when compared
to the original radial grid coordinate of the \texttt{rns}-code. For the angular grid
coordinate we choose between $t =\cos(\theta)$ and $t = 1 - 2\,\theta/\pi$ depending on the degree of differential rotation. In the Cowling
approximation, only the interior of the star needs to be considered and this
means that $s\leq 0.5$ for all $t$. In the non-rotating limit, the surface of
the star coincides exactly with the gridline at $s = 0.5$.

However, this choice of grid parameters poses a small challenge when applying
boundary conditions on the surface of a rapidly rotating star, since in this
case the surface lies in between the grid points of our computational domain.
One therefore has to approximate the outer boundary in a proper way which is
done here in the following manner: For each angular direction, we monitor the
variation of the energy density when moving radially outwards and set the
surface at this particular angle to be the first grid point where the energy
density vanishes. The boundary conditions are then applied to these set of
grid points. Depending on the shape of the star, one may have to increase the
angular resolution of the grid in order to get an accurate approximation of
the stellar surface. As already mentioned before, we use two different angular
grid coordinates. While the mapping $t = \cos(\theta)$ is appropriate for small
degrees of differential rotation, for higher degrees $t = 1 - 2\,\theta/\pi$ is the better choice.

Concerning the time-evolution scheme, we use standard central differences for
spatial discretization and the classical fourth-order Runge-Kutta method as time
integrator. Similar to previous studies within the same perturbative framework
\cite{Gaertig:2009rm,PhysRevD.78.064063}, this code is prone to numerical
instabilities around the origin of the star, which can be cured by using
standard second-order Kreiss-Oliger dissipation \cite{KreissOliger199602}.
Depending on the resolution
of the grid, we always use the smallest amount of artificial viscosity in
order to keep the effect of this purely numerical additive as marginal as
possible. Typically, the simulations were performed using a grid size of $120
\times 90$ points; in this case, the dissipation coefficients are of the order
of $10^{-4}$. As it turned out, the mode frequencies do not change
significantly when increasing either the resolution of the computational
domain or the amount of artificial viscosity. Additionally, the utilization of
the Runge-Kutta method as time-integrator allowed for a further decrease in
the amount of dissipation when compared to the Iterated Crank-Nicholson scheme
used in previous studies. This also leads to a considerably slower numerical
damping of the oscillations.

The code is numerically stable for simulations up to some hundreds of milliseconds.
Usually, our simulations cover roughly $100\unit{ms}$, resulting in a frequency
resolution of $\Delta f\approx10\unit{Hz}$. We then extract the
eigenfrequencies from the time evolution by perfoming a Fast Fourier Transform on the obtained time series at some arbitrary points inside the star. For
an unambiguous identification of a specific mode, we use Fourier transforms at
each single point inside the star in order to construct the entire
two-dimensional eigenfunction. Once we determined an eigenfrequency by means
of the Fourier spectrum at a certain point, we track the amplitude of the
Fourier transform for all grid points at this particular frequency which gives us the modulus of the corresponding eigenfunction. A sign
flip in the eigenfuction is reflected by a shift of $\pi$ in the Fourier
transform's argument. With the eigenfunctions extracted, we
are also able to perform mode recycling \cite{Dimmelmeier:2005zk,
PhysRevD.78.064063} in order to enhance a particular mode in the
oscillation pattern, typically resulting in a more accurate frequency determination.

\section{Results}
\label{sec:results}
We now present the results obtained with our method as outlined in the previous sections. First, we will test it on uniformly rotating models to check the accuracy of the newly developed code. We then turn to differentially rotating stars and investigate the effects for both small and large degrees of differential rotation on the onset of the CFS unstable fundamental quadrupolar mode. Finally, we will take a look at the generically unstable r-mode in differentially rotating stars.

\subsection{Uniform Rotation}
\label{sec:results_uniform}

We first want to consider perturbations of uniformly rotating stars mostly for code verification purposes. Barotropic oscillations of this type have already been studied extensively in
\cite{PhysRevD.78.064063} with the very same polytropic equations of state considered in this study. However, here we use a completely different set of perturbation equations, different coordinate systems and numerical methods. Still, both axisymmetric and non-axisymmetric mode frequencies
can be reproduced with a very good accuracy; the discrepancy between the two
different codes is below 2\% in all cases.

\begin{figure}[ht!]
    \centering
    \includegraphics[width=0.5\textwidth]{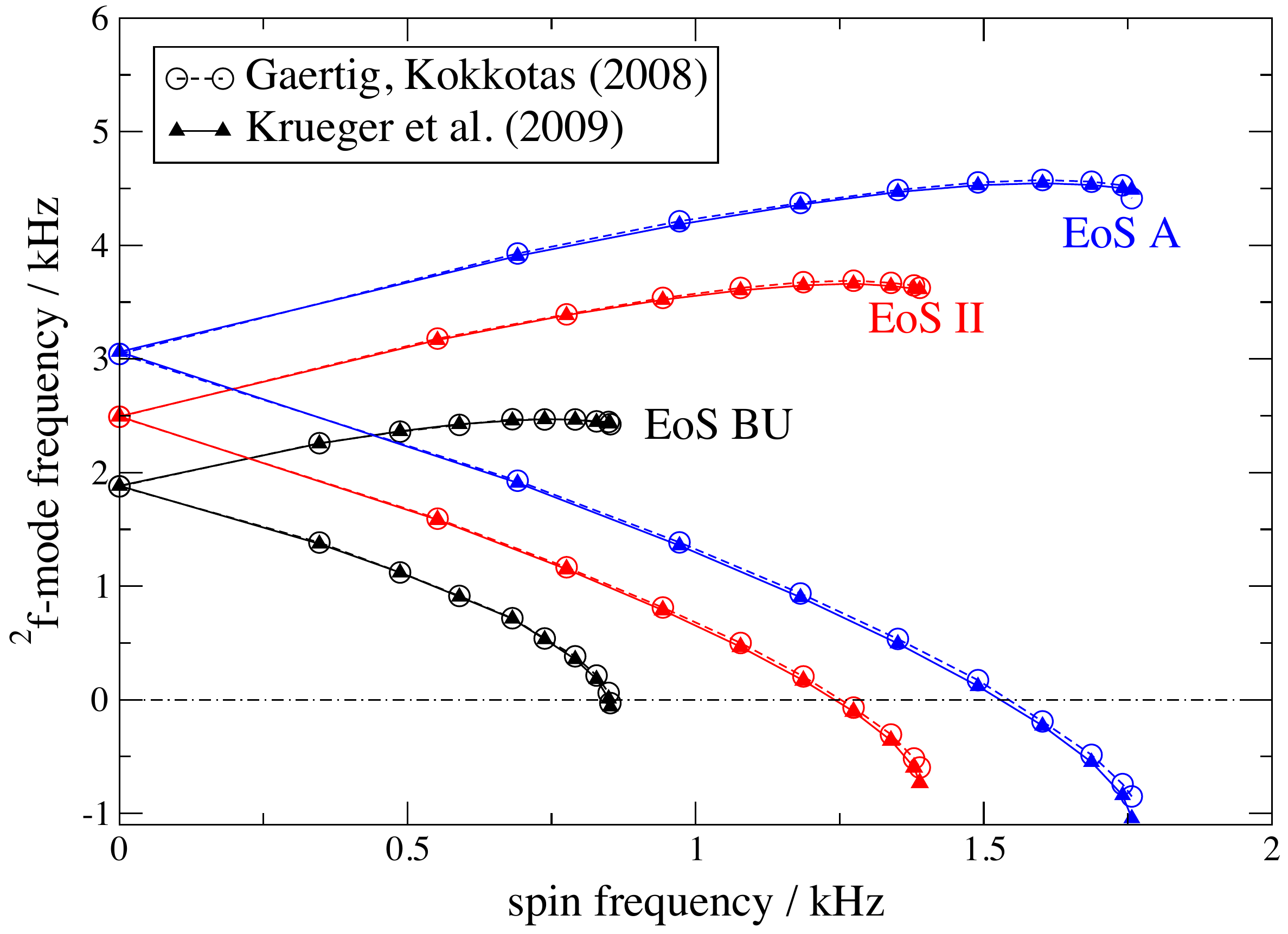}
    \caption{Comparison of the fundamental mode splitting for $l = |m| = 2$ and different EoS. Differences can be monitored by the shift of the triangles (present study) with respect to its enclosing circles (results from \cite{PhysRevD.78.064063}).}
    \label{fig:comparison}
\end{figure}

As an example, Figure \ref{fig:comparison} shows a comparison between results obtained in the present study and published values for the fundamental quadrupolar f-mode splitting found in \cite{PhysRevD.78.064063}. The sequences of background models in these two studies are identical, so a good means in order to estimate the differences between the two methods is to observe the off-centering of the triangles, which are the data points from the current study, from its enclosing circles, which represent the literature values. 

Obviously, only for very high rotation rates there are some deviations; otherwise the two codes provide nearly identical results.

We also checked the frequencies of the second class of oscillation modes which
are present in a barotropic, perfect fluid star as soon as rotation sets in;
these are the inertial modes. Again, the agreement between the two codes
is excellent; a comparison with frequencies found in
\cite{2008PhRvD..77l4019K} also shows a good agreement (see \cite{Kruger:2009dq}
for more details).

\subsection{Axisymmetric Modes of Differentially Rotating Stars}
\label{sec:results_axisymmetric}

Next, we consider axisymmetric perturbations on differentially
rotating backgrounds and compare the results with \cite{Stergioulas:2003ep}, who investigated the same problem with a
non-linear code applied to small perturbations.

\begin{figure}[ht!]
    \centering
    \includegraphics[width=0.5\textwidth]{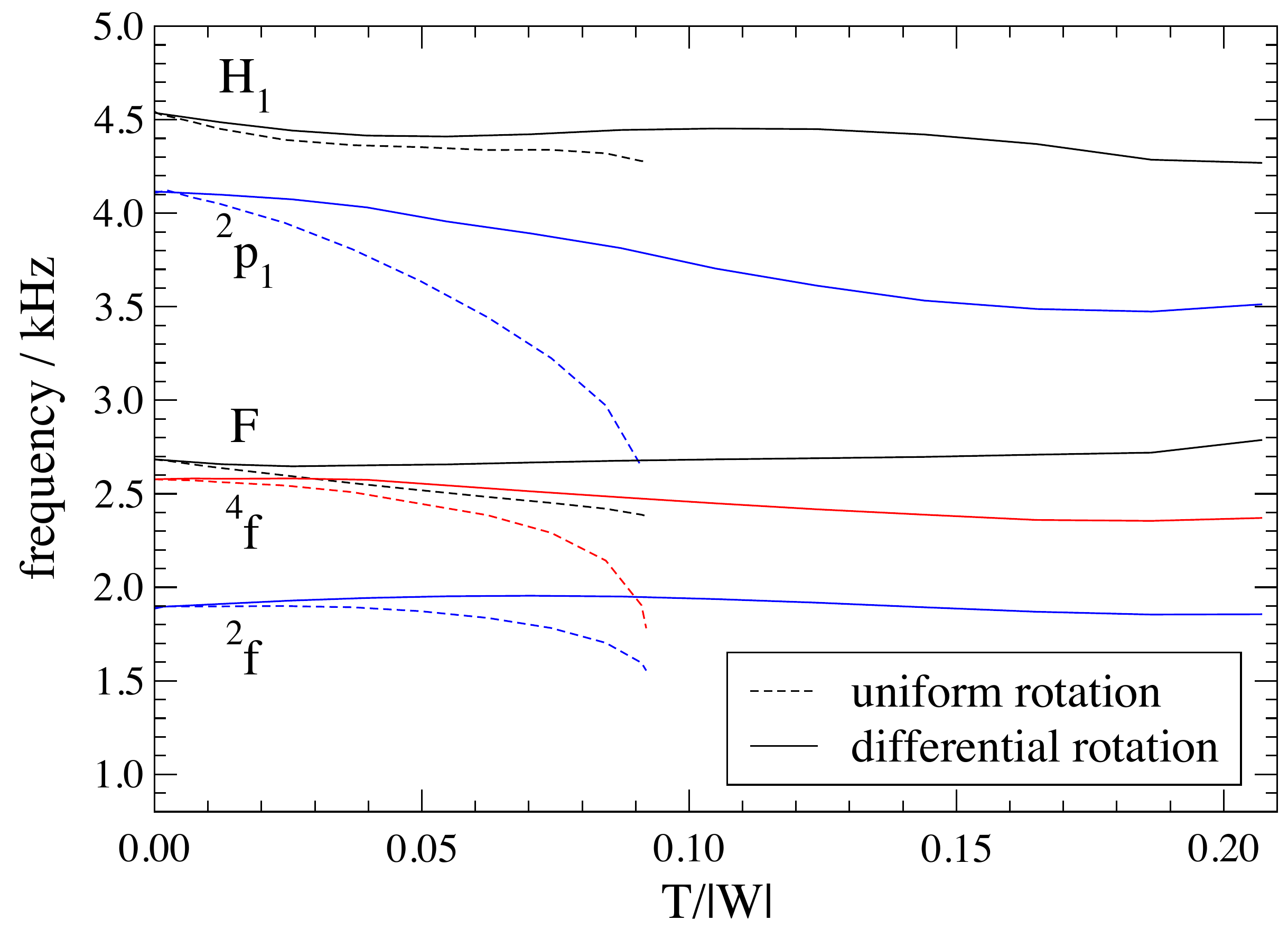}
    \caption{Mode frequencies of several quasi-radial $(m = 0)$ modes for uniformly and differentially rotating stars of the B sequence.}
    \label{fig:krueger_m0}
\end{figure}

Figure \ref{fig:krueger_m0} shows the effect of rotation on the three
fundamental modes F, ${}^2$f and ${}^4$f as well as the first overtones H${}_1$ and
${}^2$p${}_1$ for both uniformly and differentially rotating models of the BU and B sequences. As already discussed in Section \ref{ssec:equilibriumModels}, differential rotation allows for a higher mass of equilibrium models compared to their uniformly rotating counterparts and the corresponding value of
$T/|W|$ can reach significantly higher values. This is the reason why the
dashed curves for uniformly rotating stars terminate already at a rather low
$T/\left|W\right|$ in Figure \ref{fig:krueger_m0}, which is similar to Figure
5 in \cite{Stergioulas:2003ep}.

As one can clearly see, the mode frequencies of differentially rotating models
are typically larger than the corresponding frequencies for the uniformly
rotating sequence. The reason for this is that differentially rotating stars
usually have a smaller equatorial radius than their uniformly rotating counterparts
with the same
$T/\left|W\right|$. Therefore, the crossing time of acoustic waves, like the f-
and higher order p-modes, is lowered which leads to a higher oscillation frequency.

The agreement with the data provided in \cite{Stergioulas:2003ep} is very good; the
relative difference is below 2.5\% for all frequencies. However, we disagree about the interpretation of the so-called F$_{\rm II}$-mode which was found in \cite{Stergioulas:2003ep} and was interpreted as a possible splitting of the fundamental quasi-radial mode in differentially rotating stars; a reason for this might be the Cowling approximation which violates energy and momentum constraints.

When we reiterated this simulation with the code presented in this study, we
also found peaks in the Fourier spectrum at the very same frequencies
published in \cite{Stergioulas:2003ep} for the F$_{\rm II}$-mode. However,
after extracting the corresponding eigenfunction, we are convinced that this
F$_{\rm II}$-mode is nothing more than the ${}^4$f-mode. The
frequencies of both the F-mode and the ${}^4$f-mode are nearly indistinguishable in slowly
rotating stars; the difference is around $100\unit{Hz}$  which makes a reliable
assignment of the eigenfunctions a difficult task. Fortunately, it turned out that the eigenfunction of $\delta u^{\theta}$ is a very good means to discriminate between the two quasi-radial modes.

\begin{figure}[ht!]
    \centering
    \includegraphics[width=0.49\textwidth]{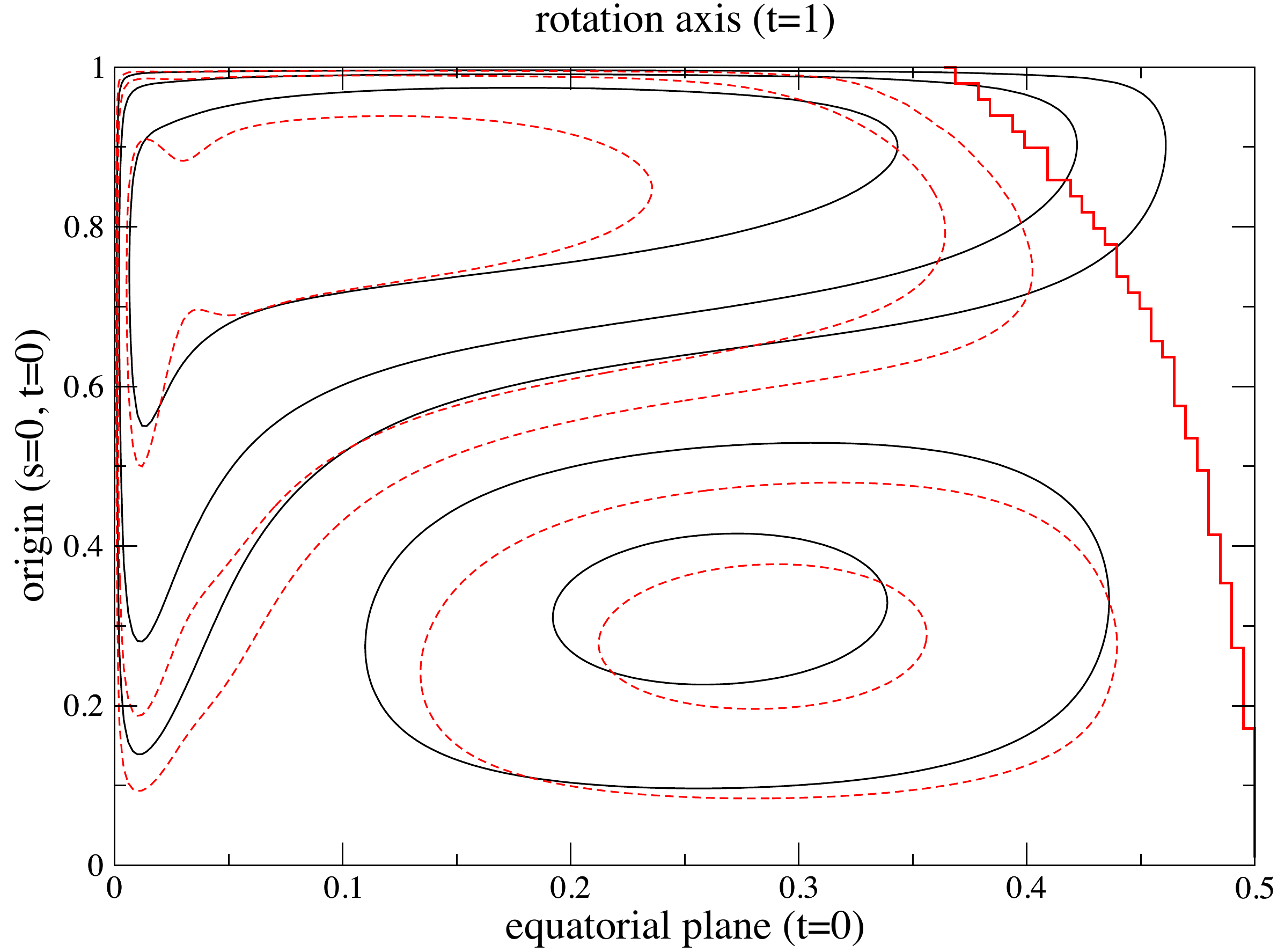}
    \includegraphics[width=0.49\textwidth]{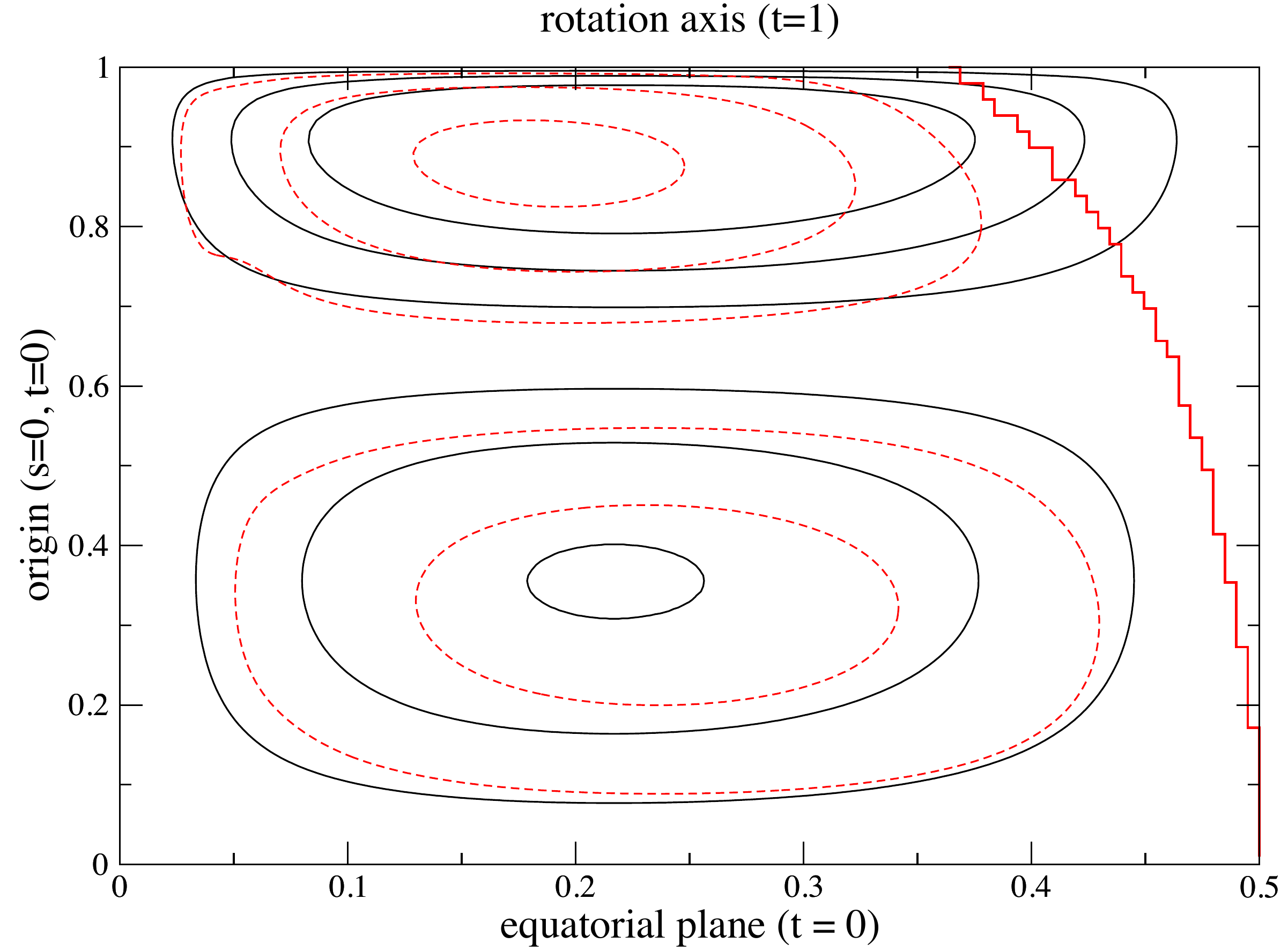}
    \caption{{\it Left Panel:} Contours of the $Q_4$-eigenfunction of the
    F-mode for a slowly rotating B model with $r_p/r_e = 0.999$ (solid) and a
    B3 neutron star (dashed). The step-like curve on the right approximates
    the stellar surface for the B3 model. {\it Right Panel:} Same for the $^4$f-mode.}
    \label{fig:contour_F_and_4f}
\end{figure}

Figure \ref{fig:contour_F_and_4f} shows contour lines of the time-evolution
variable $Q_4\sim\delta u^{\theta}$ (see section
\ref{ssec:boundaryConditions}) projected on the $(s,t)$-plane for the F- and
the $^4$f-mode. In both panels, these modes are compared for a very slowly rotating model (solid line) with an axis ratio of $r_p/r_e = 0.999$  and a B3 neutron star (dashed line). The step-like curve on the right side of both panels is the approximated surface of the B3 model. As already discussed in Section \ref{sec:numericalTreatment}, only for non-rotating stars the stellar surface coincides with $s=0.5$. Within the grid-resolution used in Figure \ref{fig:contour_F_and_4f}, this is still true for the slowly rotating model considered here. The rotation rate of the B3 model is also the reason, why the contours of $Q_4$ are shifted to smaller radial distances in this case. Apart from that, the shapes of the eigenfunctions are identical in both panels.

Actually, this issue about erroneous mode identification will always be present for various stellar models and certain modes. In the case shown here, the problem can be resolved if one drops the Cowling-approximation because this approximation overestimates the frequencies of the F-modes, making them nearly identical to those of the $^4f$-mode for this particular sequence of equilibrium models. However, the problem will still be there for different modes or stellar models and thus the analysis presented here is the only meaningful way to distinguish between certain modes.

\subsection{Non-Axisymmetric f-modes in Differentially Rotating Stars}
\label{ssec:results_nonaxisymmetric}

We next turn to non-axisymmetric perturbations on differentially rotating
stars. In the following discussion, we will mostly focus on the quadrupolar fundamental mode in
particular, because this is the lowest order mode which is able to emit gravitational waves. For uniformly rotating
stars, the ${}^2$f-mode is also prone to the secular CFS instability, but
typically only for models which rotate close to their mass-shedding limit. The
neutral point for the onset of the instability is considerably lower for modes with higher azimuthal index $m$. However, it turns out that the growth time of these modes is considerably larger than the viscous timescales. Therefore, one usually studies the $^2$f-mode exclusively.

In Table \ref{tab:freq_quadrupolar_B} we show the eigenfrequencies of the $^2$f-mode for the sequence B and a degree of differential rotation of $\hat{A} = 1.0$. As one can see, the star becomes unstable at a point between the B8 and the B9 model. The values in Table \ref{tab:freq_quadrupolar_B} have been compared with preliminary results from non-linear evolution codes and show a good agreement \cite{Zink:yq}.

\begin{table}[ht!]
  \centering
  \begin{tabular}{ccc}
    \hline
    model & \multicolumn{2}{c}{ ${}^2$f (kHz) }  \\
           & ~~$m=2$~~  & ~~$m=-2$~~  \\
    \hline
    B0    &  1.884 &  1.884 \\
    B1    &  1.428 &  2.239 \\
    B2    &  1.204 &  2.363 \\
    B3    &  1.015 &  2.449 \\
    B4    &  0.841 &  2.515 \\
    B5    &  0.673 &  2.567 \\
    B6    &  0.505 &  2.611 \\
    B7    &  0.338 &  2.648 \\
    B8    &  0.162 &  2.683 \\
    B9    & -0.026 &  2.726 \\
    B10   & -0.246 &  2.788 \\
    \hline
  \end{tabular}
   \caption{Frequencies of the $^2$f-mode for the B sequence and $\hat{A} = 1.0$.}
  \label{tab:freq_quadrupolar_B}
\end{table}

One may now ask, how this behaviour changes for different degrees of
differential rotation. As we know, the uniformly rotating model, which
corresponds to $\hat{A}^{-1} = 0.0$, becomes secularly unstable just around its
mass-shedding limit while for $\hat{A}^{-1} = 1.0$ this happens at roughly
65\% of the highest possible $T/|W|$ for a stable equilibrium.

\begin{figure}[ht!]
    \centering
    \includegraphics[width=0.49\textwidth]{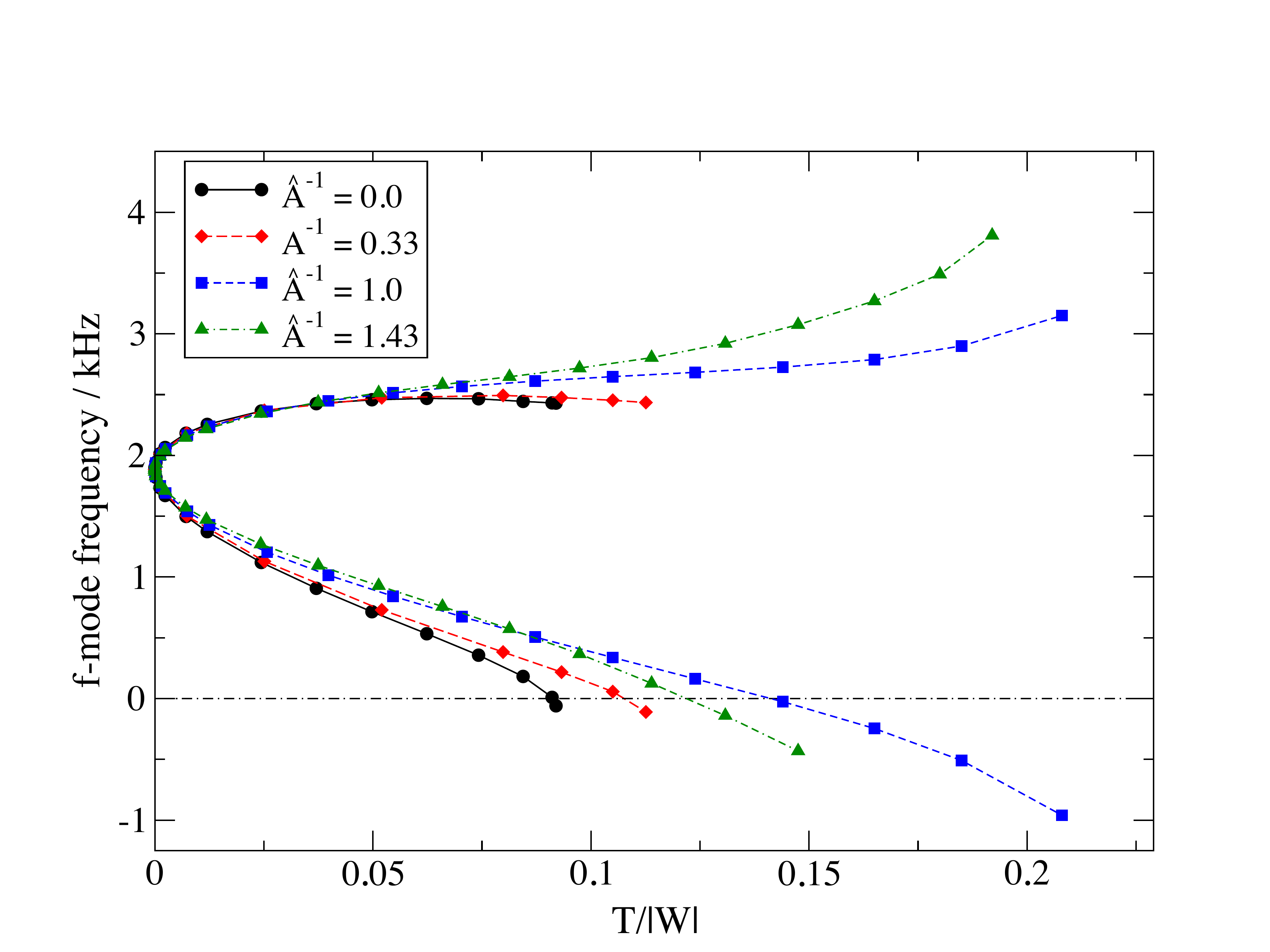}
    \includegraphics[width=0.5\textwidth]{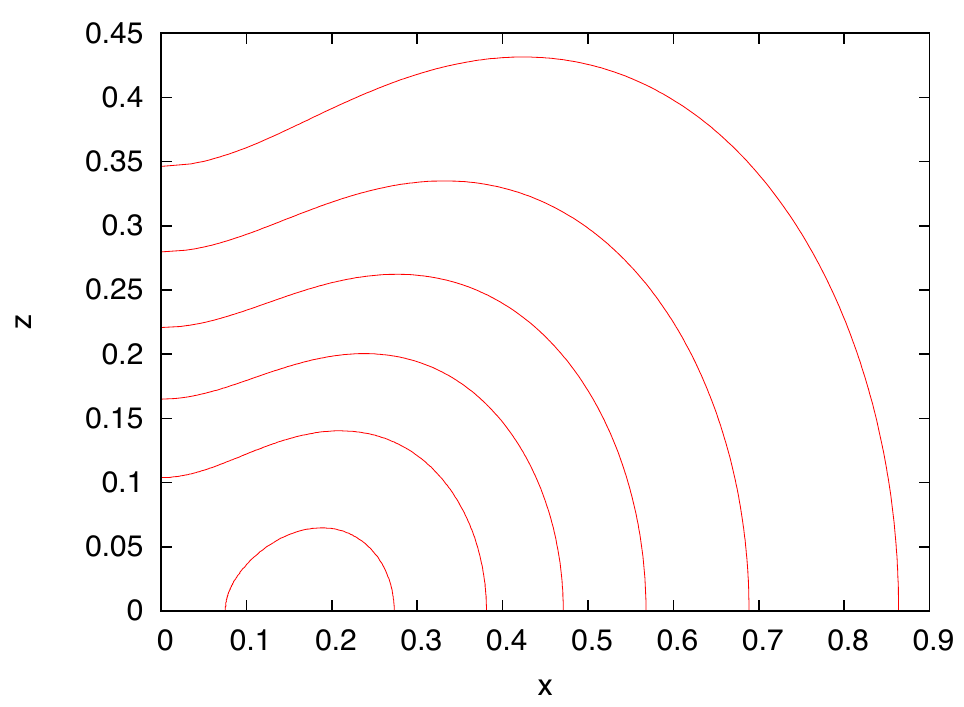}
    \caption{{\it Left Panel:} Splitting of the ${}^2$f-mode for different degrees of
    differential rotation $\hat{A}$. All equilibrium sequences are based on EoS B. {\it Right Panel:} Density contour lines of the most rapidly rotating model B12. The outermost line corresponds to the surface of the star, the z-axis represents the rotation axis.}
    \label{fig:2f_splitting_B}
\end{figure}

Figure \ref{fig:2f_splitting_B} shows the splitting of ${}^2$f-mode for four different degrees of
differential rotation, starting from the uniform rotation limit with $\hat{A}^{-1} = 0.0$ to a value of $\hat{A}^{-1} = 1.43$. First of all, one can see that differential rotation leads to an increase in the mode frequencies compared to the uniformly rotating case; something we already observed for axisymmetric modes, see Figure \ref{fig:krueger_m0}. Second, a high degree of differential rotation favours the onset of the CFS instability in the sense that the neutral point is reached at a smaller fraction of the maximum allowed $T/|W|$.

Third, for sufficiently large values of $\hat{A}^{-1}$, the mode frequencies of the corotating branch always increase as can be seen from Figure \ref{fig:2f_splitting_B} for $\hat{A}^{-1}\gtrsim 1.0$. This is in strong contrast to the uniformly rotating case where the frequencies of the prograde travelling modes reach a local maximum before they decrease again; see also Figure \ref{fig:krueger_m0}. If the background models allowed for high rotation rates, the frequency of the corotating branch would decrease even more and at some value of $T/|W|$ merge with the frequency of the retrograde travelling mode which is dragged forward by rotation in the inertial frame. This point then signals the onset of the dynamical bar-mode instability.

The situation changes in the case of differential rotation. Depending on the degree of differential rotation and the rotation rate, there is a transition of the axisymmetric equilibrium configuration from an oblate spheroid with its maximum energy-density and pressure at the center of the star towards a toroidal configuration where these peak values are reached along a ring-like structure in the equatorial plane; see the right panel of Figure \ref{fig:2f_splitting_B} for a density-contour plot of the most rapidly rotating model with $\hat{A}^{-1} = 1.0$. This change in the topology of the background model leads to an increase in the frequencies of the corotating branch which of course sets in earlier for equilibrium configurations with large degrees of differential rotation since in this case, the toroidal structure of the background star is already realized for comparatively small values of $T/|W|$. One should also note that while the sequences considered here are still configurations with the same central energy density, the maximum value of the energy density is reached outside the center of the star as soon as the transition to a toroidal structure  sets in.

\subsubsection{Small Degree of Differential Rotation}

We want to discuss the onset of the CFS-instability a bit more in the regime of weakly
differential rotation, i.\,e. small values of  $\hat{A}^{-1}$. As one can see clearly from Figure \ref{fig:2f_splitting_B}, the critical value of $\beta:=T/|W|$, where the neutral point of the secular instability is reached, typically increases with the degree of differential rotation. In order to estimate whether differential rotation favours the onset of the CFS mechanism, one also needs to know the maximum value $\beta$ can attain. In differentially rotating stars, a large amount of angular momentum can be stored in inner layers close to the rotational axis. This also means, that the highest possible value of $\beta$ increases as well.

In the subsequent discussion, we will restrict our simulations to smaller
values of $\hat{A}^{-1}$ for the following reason: Since we later want to
normalize the critical value for the onset of the CFS instability $\beta_c$
with the corresponding value $\beta_s$ at the mass-shedding limit, we need
dynamically stable equilibrium models at this Kepler frequency; otherwise the
normalization would not make much sense. It turns out, that our particular
choice of background models is dynamically unstable to radial oscillations for comparatively small
values of $\hat{A}^{-1}$. For example, the B sequence is dynamically unstable
before reaching the mass-shedding limit for $\hat{A}^{-1}\gtrsim 0.7$. It is
worth pointing out, that the endpoint of the B model series in Table
\ref{tb:equilibriumModels}, i.\,e.~model B12, does not terminate because of the mass-shedding limit but because of dynamical instabilities. The same is also true for sequences A and II; due to the higher initial masses, this effect is even more pronounced there.

\begin{figure}[ht!]
    \centering
    \includegraphics[width=0.49\textwidth]{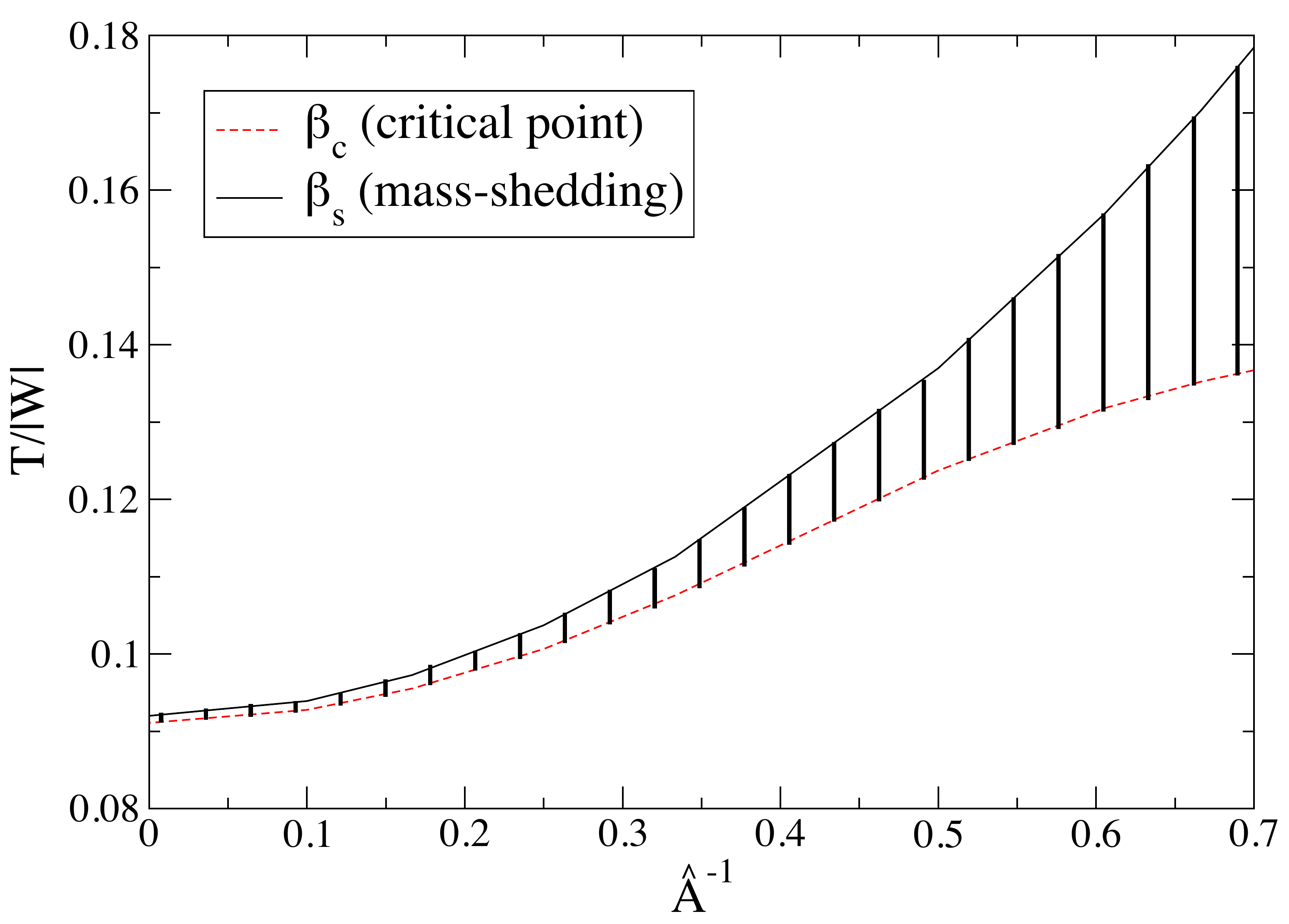}
    \caption{The values of $\beta_c$ and $\beta_s$ as a function of
    $\hat{A}^{-1}$ for the $^2$f-mode and the B model series. The filled area marks the CFS-unstable region.}
    \label{fig:tw_crit_kepler_B}
\end{figure}

Figure  \ref{fig:tw_crit_kepler_B} shows, how $\beta_c$ and $\beta_s$ change with the degree of differential rotation. As already conjectured, both values increase with $\hat{A}^{-1}$ and the slope of $\beta_s$ is even larger than the one for $\beta_c$. This confirms our initial assumption, that the onset of the secular CFS instability is eased for larger degrees of differential rotation.

\begin{figure}[ht!]
    \centering
    \includegraphics[width=0.5\textwidth]{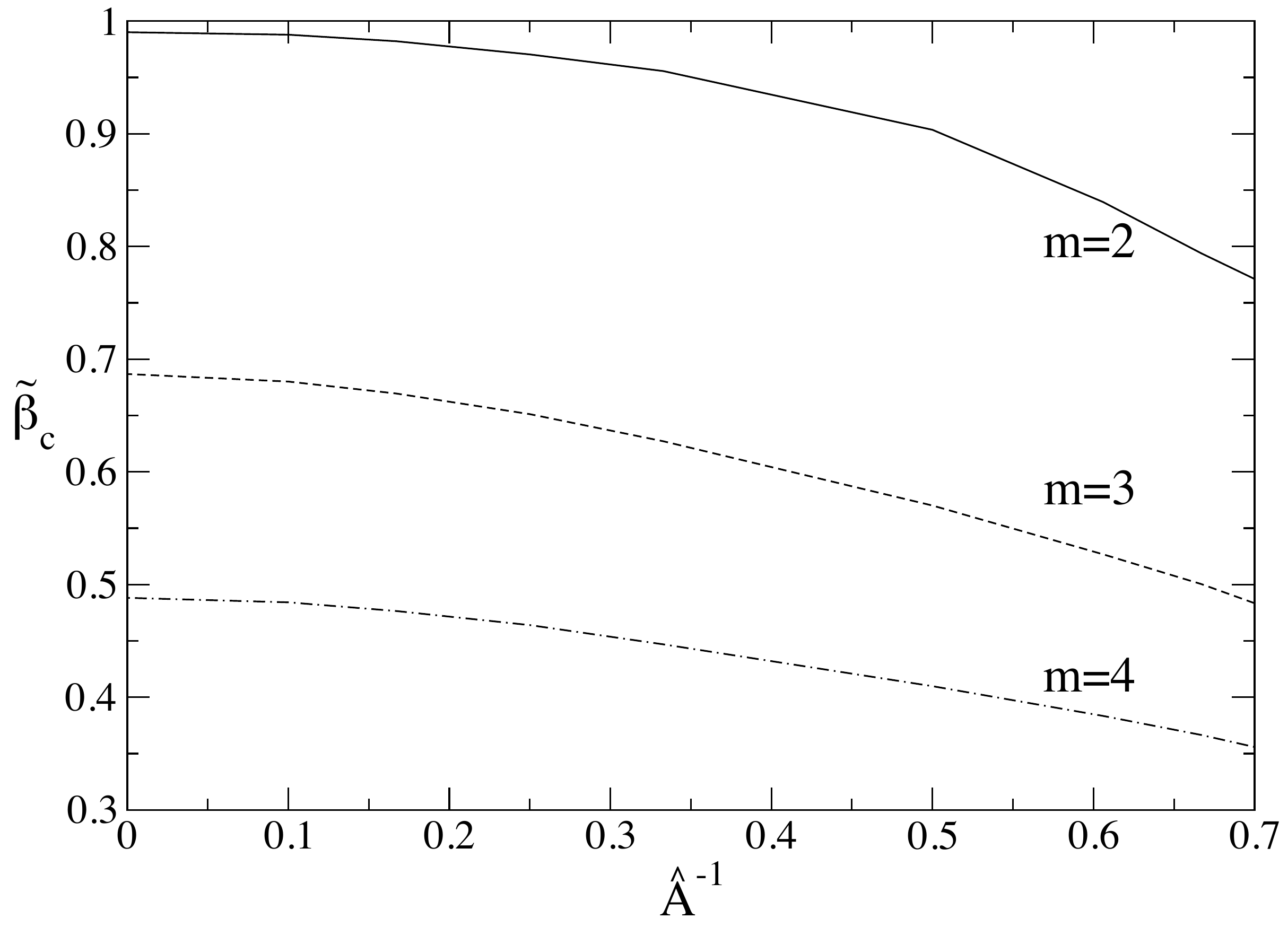}
    \caption{The normalized, critical value $\tilde{\beta_c}$ for the onset of
    the CFS instability of the B sequences as a function of the degree of
    differential rotation $\hat{A}^{-1}$ for $m = 2,3,4$.}
    \label{fig:tilde_beta_c_over_A_m234}
\end{figure}

In Figure \ref{fig:tilde_beta_c_over_A_m234}, we plot the value of $\tilde{\beta}_c:=\beta_c/\beta_s$ for three different modes of the B sequence as function of the degree of differential rotation $\hat{A}^{-1}$. This normalization of the critical $T/|W|$ is similar to the common procedure for uniformly rotating stars, where one typically measures the angular velocity of the star in units of the Kepler-frequency.

Keep in mind that the uniformly rotating limit is reached for $\hat{A}^{-1} =
0.0$. As we already have seen in Figure \ref{fig:2f_splitting_B}, in this case
the $^2$f-mode becomes unstable just around its mass-shedding limit which
corresponds to $\tilde{\beta_c}\approx 1.0$. It has already been mentioned in
Section \ref{ssec:results_nonaxisymmetric} that higher order modes become
secularly unstable even earlier and this is also reflected in Figure
\ref{fig:tilde_beta_c_over_A_m234}. For example, in the limit of uniform rotation,
the $^3f$-mode gets unstable at roughly 70\% of the critical $T/|W|$, the
$^4$f-mode even at approximately 50\%. However, it has already been noted that the growth time for the instability will increase for modes with larger $m$ and is most likely suppressed by viscosity.

When we trace the value of  $\tilde{\beta}_c$ for the various fundamental
modes and several degrees of differential rotation, we see clearly that for
larger values of $\hat{A}^{-1}$, the critical $T/|W|$ decreases invariably for
all modes; i.\,e. the onset of the instability is favoured in these cases.

\begin{figure}[ht!]
    \centering
    \includegraphics[width=0.5\textwidth]{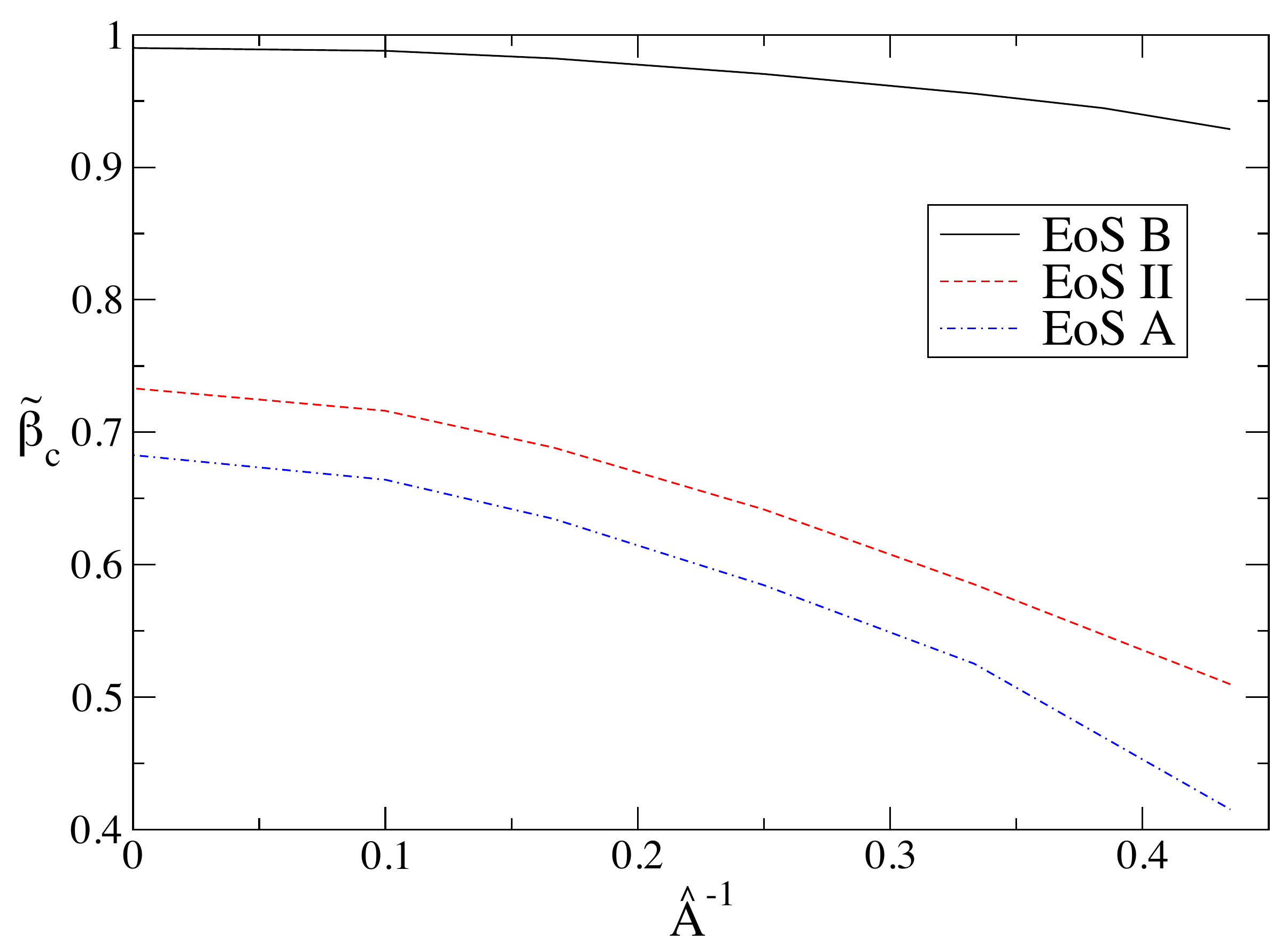}
    \caption{The normalized, critical value $\tilde{\beta_c}$ for the onset of
    the CFS instability of the ${}^2$f-mode for different equations of state.}
    \label{fig:tilde_beta_c_over_A_eos}
\end{figure}

We also studied the behaviour of $\tilde{\beta}_c$ for the stiffer equations of state EoS A and EoS II. The actual equilibrium models we
considered are very close to their maximum allowed mass and therefore we can
increase the differential rotation parameter $\hat{A}$ only to a very moderate
degree of around $\hat{A}\approx 2.3$ because models with a higher degree of
differential rotation will be dynamically unstable. Figure
\ref{fig:tilde_beta_c_over_A_eos} again shows a plot of $\tilde{\beta}_c$ for
different values of $\hat{A}^{-1}$, but this time for the fundamental
quadrupolar mode and the three different polytropic equations of state that
were considered in this study. As already found in \cite{PhysRevD.78.064063}
and one can also see in Figure \ref{fig:comparison}, the CFS instability acts
much earlier in the more compact sequences A and II.

It is worth noticing that the influence of differential rotation is even more
pronounced for these sequences when compared to sequence B. As one can see
from Figure \ref{fig:tilde_beta_c_over_A_eos}, the critical curves for EoS A
and EoS II decrease more strongly than the corresponding curve for the less
compact EoS B series. As an example, while increasing the differential
rotation parameter from $\hat{A}^{-1}= 0.0$ to $\hat{A}^{-1}= 0.43$, the value
of $\tilde{\beta}_c$ is lowered only by 6\% for sequence B but by 30\% for
sequence II and even 39\% for sequence A. Therefore, the onset of the
CFS instability is also eased by the compactness of the neutron star when
differential rotation is considered.

\subsubsection{Large Degree of Differential Rotation}

As we discussed in the previous section, equilibrium sequences with a high
degree of differential rotation ($\hat{A}^{-1} \gtrsim 1$) terminate due to
dynamical instabilities with respect to radial oscillations; see \cite{Baumgarte:2000rr} for a study on the maximum mass of differentially rotating neutron stars. Therefore, it
does not make sense to define the mass-shedding value $\beta_s$, and
correspondingly the same holds for the normalized quantity $\tilde{\beta}_c$.
Nevertheless, the sequences still reach the neutral point $\beta_c$ for the
onset of the CFS instability of the fundamental quadrupolar mode.

\begin{figure}[ht!]
    \centering
    \includegraphics[width=0.47\textwidth]{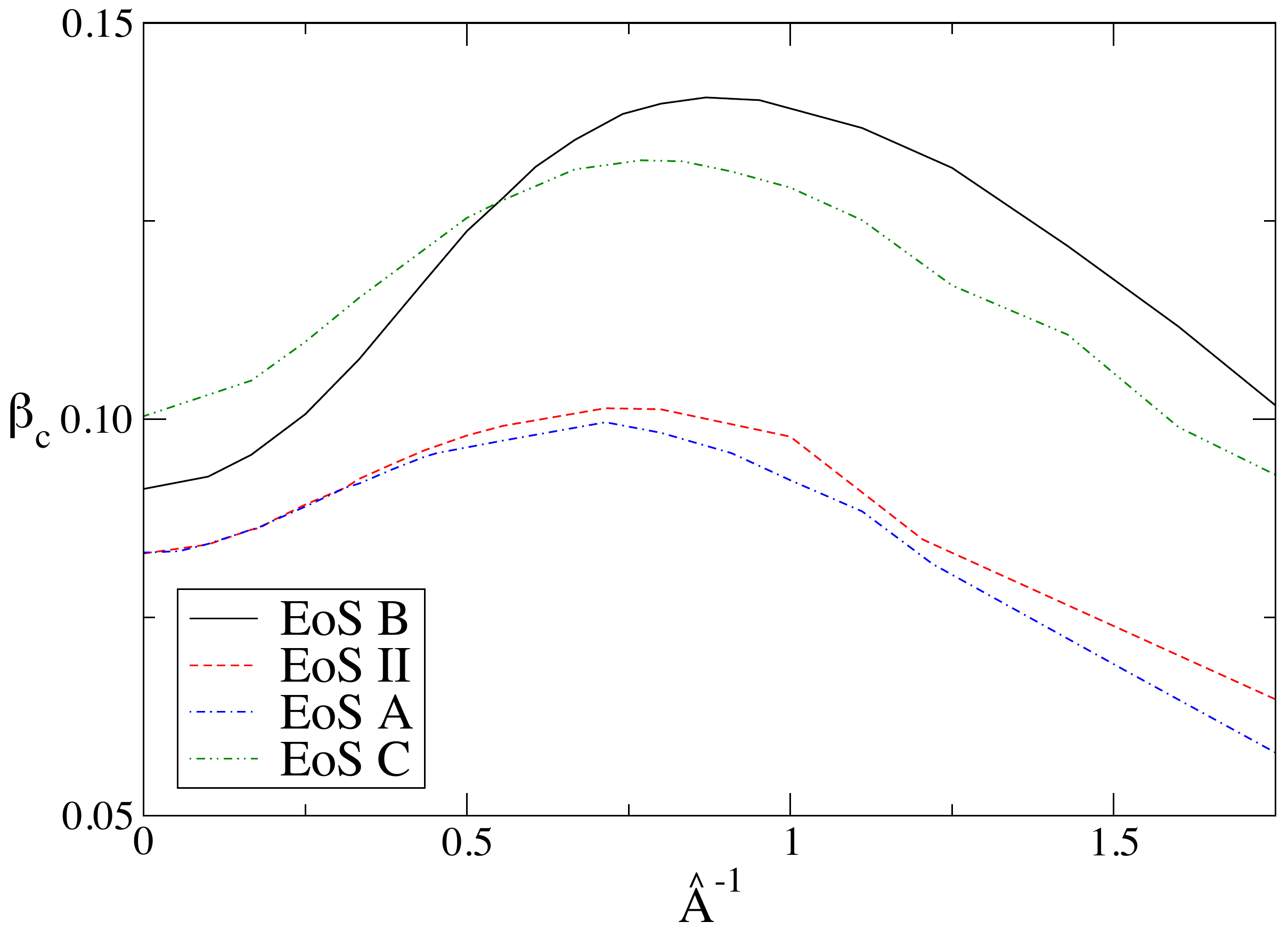}
    \caption{The critical value $\beta_c$ of the ${}^2$f-mode for four
    different equations of state.}
    \label{fig:critical_curve_high_diff}
\end{figure}

In Figure \ref{fig:critical_curve_high_diff}, we show $\beta_c$ of the
quadrupolar f-mode for the three equations of state EoS B, EoS II and EoS A as well as for an additional one which we label EoS C and which has an intermediate polytropic exponent of $\Gamma = 2.18$ and $K = 400$ in the same units as for the other equations of state.

In contrast to the previous section, we continue to increase the degree of
differential rotation up to $\hat{A}^{-1} = 1.75$. When moving from rigid rotation to a small degree of differential rotation, the value of $\beta_c$ increases for each equation of state. Any sequence considered in this study reaches a local maximum which is located at $\hat{A}^{-1} = 0.89$ for the B-sequence and $\hat{A}^{-1} = 0.78$ for EoS C while EoS II and EoS A reach it considerably earlier at $\hat{A}^{-1} = 0.74$ and $\hat{A}^{-1} = 0.69$,
respectively.

For higher degrees of differential rotation, the value of $T/|W|$ for the
neutral point decreases. Again, similar to the observations we made for small degrees of differential rotation, this effect is more pronounced for the more compact models of the stiffer equations of state EoS II and EoS A. As one also can see from Figure
\ref{fig:critical_curve_high_diff}, the critical value for the C, A and II
sequences even drop clearly below the value for rigid rotation. For Eos A and EoS II, this happens roughly at around $\hat{A}^{-1} \approx 1.2$ while $\hat{A}^{-1} \approx 1.5$ for EoS C. In contrast to this, while also decreasing in amplitude, the value of $\beta_c$ for the B-sequence does not drop below the corresponding value of the uniform rotation limit, at least not up to  $\hat{A}^{-1} = 1.75$ which is the endpoint of the sequences in Figure \ref{fig:critical_curve_high_diff}.

A similar investigation has already been carried out for Newtonian polytropes
\cite{2002ApJ...578..413K}, and we qualitatively agree with their results.
Further comparisons are not possible for various reasons; most crucial is
the different definition of the parameter $\hat{A}$ which controls the degree of
differential rotation.

\subsection{Rotational Modes in Differentially Rotating Stars}
\label{ssec:rotational_modes}

Another very important class of oscillation modes are the
r-modes, because they are generically CFS unstable
\cite{1998ApJ...502..708A}. Here, we only consider the $l=m=2$ r-mode. Since
the r-mode is purely axial in the non-rotating limit, it belongs to that
class of modes whose pressure variation has odd parity under reflection with
respect to the equatorial plane.

\begin{figure}[ht!]
    \centering
    \includegraphics[width=0.5\textwidth]{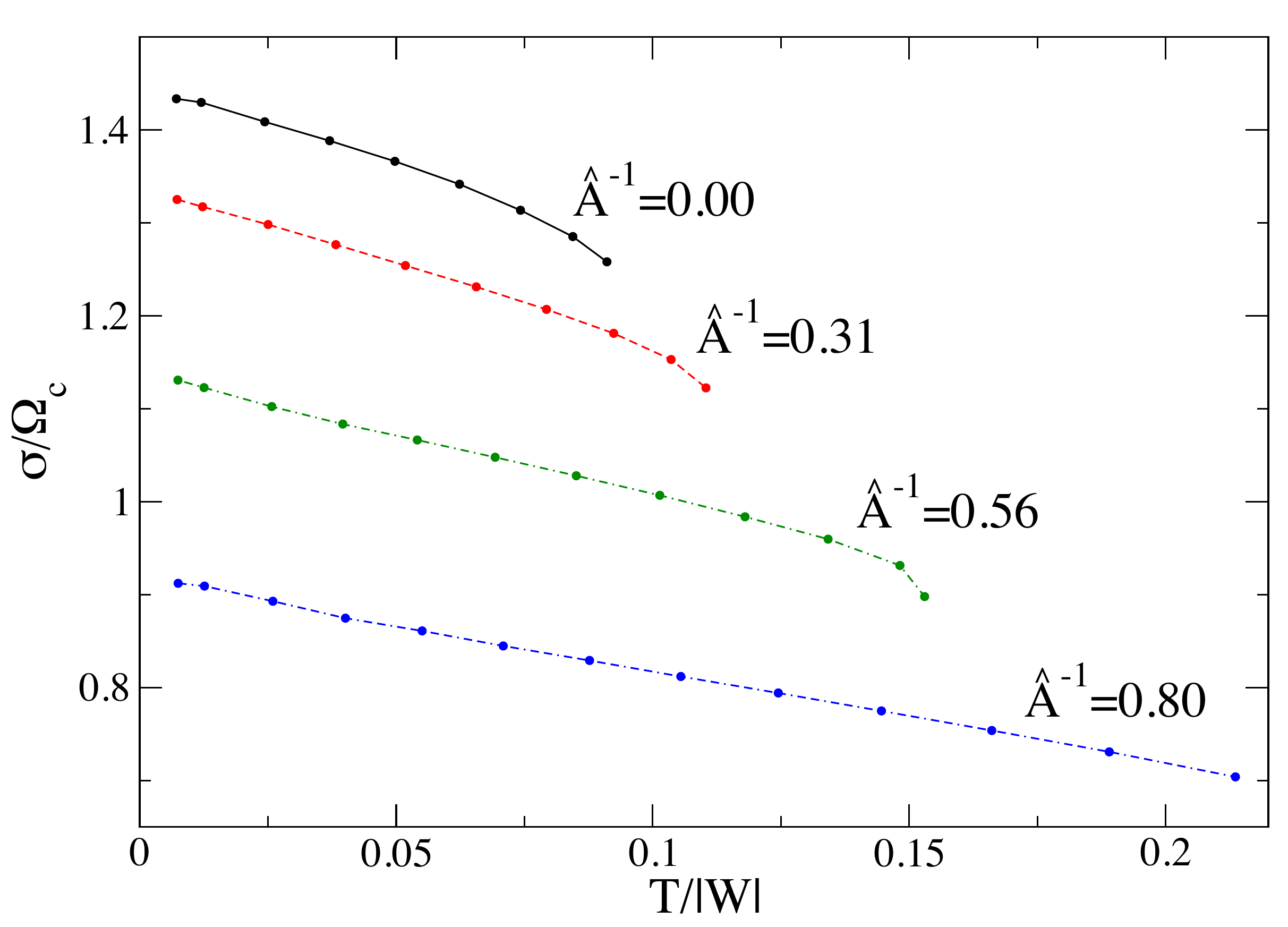}
    \caption{Frequencies of the r-mode for several B sequences with
    different degrees of differential rotation. The frequencies are normalized
    by the central angular velocity $\Omega_c$.}
    \label{fig:r_mode}
\end{figure}

In Figure \ref{fig:r_mode} we show the frequencies of the r-mode along
sequences using EoS B of for different degrees of differential rotation. We
normalize them by the central angular velocity $\Omega_c$ of the corresponding
star, as it has already been done in \cite{PhysRevD.64.024003}. This kind of
normalization is suggested by the relation $\sigma = 4/3\, \Omega$
which is valid in Newtonian theory, where $\sigma$ is the r-mode's
angular frequency measured in the inertial frame and $\Omega$ the angular
velocity of the star.

First, we inspect the solid curve representing the rigidly rotation case,
i.\,e. $\hat{A}^{-1}=0$. Concerning the value of $\sigma/\Omega_c$ in the
non-rotating limit, which is approximately 1.43, we observe a clear deviation
from Newtonian theory. However, this is consistent with general relativistic
corrections to the Newtonian result. It can be shown \cite{2005MNRAS.356..217Y}
that in the non-rotating limit the Newtonian value of 4/3 is increased by a term which is proportional to the frame-dragging potential $\omega$.

The curves corresponding to the differentially rotating sequences have lower
values, because the central angular velocity increases compared to the uniformly
rotating ones, and this effect becomes even stronger the higher the degree of
differential rotation is. On the other hand, the angular velocity at the
equator, $\Omega_e$, is lowered so that a normalization by this quantity would
move the lines in the other direction. Finally, we chose to normalize by the
central angular velocity in order to compare qualitatively with Newtonian
findings \cite{PhysRevD.64.024003}. We find that the qualitative behaviour is
the same.

Moreover, we explicitly computed the eigenfunction of the $\delta u^{\theta}$ velocity
perturbation. According to the Newtonian results in
\cite{PhysRevD.64.024003}, we find that the amplitude of the eigenfunction
gets confined towards the surface of the star as differential rotation is
increased.

Earlier studies based on the slow rotation approximation also suggested that there might be a continuous spectrum
for the r-modes
\cite{1998MNRAS.293...49K,1999MNRAS.308..745B,2001MNRAS.328..678R}. More
recently the same claim has been made in the study of differentially
rotating relativistic stars in the slow-rotation approximation
\cite{2008PhRvD..77b4029P}. Here, in agreement with earlier results for
uniformly rotating stars \cite{PhysRevD.78.064063,Gaertig:2009rm}, we have
found no trace of the continuum spectrum. This result does not exclude its
existence, because the continuum can co-exist with discrete modes.
Especially here we excite the oscillation by using  an approximate
eigenfunction of the mode under study, then by repeated  mode recycling we
force the star to oscillate on this specific mode alone in order to
extract both the eigenfrequency and the eigenfunction with highest
possible accuracy. This way we exclude excitation of any other mode or
continuum, still we should stress that no numerical calculation of fast
rotating relativistic star modes to date both in linear and non-linear
form has shown the presence of a continuous spectrum.

\section{Summary}
\label{sec:summary}
In this work we performed time evolutions of oscillations of differentially
and rapidly rotating,
relativistic stars using the Cowling approximation, and we were able to investigate both axisymmetric and non-axisymmetric perturbations. We extracted
eigenfrequencies and eigenfunctions by applying Fourier transforms on the
obtained time-series. The agreement with literature values is excellent for
both non-axisymmetric perturbations of uniformly rotating stars and
axisymmetric perturbations of differentially rotating stars.

Our simulations show that a small degree of differential rotation causes the
ratio of the critical $T/|W|$, where the f-mode becomes CFS unstable,
to the value of $T/|W|$ at the mass-sheeding to decrease. This effect is visible
for each considered equation of state and is even stronger for the stiffer
ones. It implies that neutron stars can be destabilized by the CFS
instability even if their spinning frequency is rather low compared to the
Kepler frequency. Furthermore, when moving on to a high degree of differential
rotation, the critical curve of any considered sequence exhibits a local
maximum which is in agreement with Newtonian findings. Afterwards, the
critical value decreases and for the sequences based on the stiffer equations
of state it decreases even below the corresponding value for rigid rotation;
i.\,e. the CFS instability may act even earlier than in rigidly rotating stars
as long as the star rotates with a sufficiently high degree of differential
rotation.

Summarized, we showed that both a small and a high degree of differential
rotation plays an important role in the evolution of a neutron star.
As explained in the introduction, neutron stars rotate differentially most
probably directly after their birth, but may do so also at later times.
Whenever differential rotation is present, the CFS instability is a promising
candidate for the destabilization of a neutron star which may finally lead to
a detectable signal of gravitational waves.

Finally, we performed simulations on r-modes. Our results show a
good qualitative agreement to Newtonian results. As in earlier studies which
do not rely on the slow rotation approximation, we do not see any trace of a
continuous spectrum.

A very important issue related to differential rotation is the freedom to construct neutron star models with much higher values of $T/|W|$. This means that nearly all available equations of state can admit models which can be CFS-unstable which is not the case for uniformly rotating stars. This enhances the probability of having the CFS-instability working in the early phases of the proto-neutron star creation which will be a significant boost in our efforts to detect gravitational waves from isolated neutron stars and via such observations to infer and constrain fundamental neutron star parameters.

\section{Acknowledgments}
\label{sec:acknowledgements}

It is a pleasure to thank N.~Stergioulas for providing his \texttt{rns} code
for differentially rotating background models and for carefully reading this manuscript, providing valuable suggestions for improving the article. We also thank M.~Vavoulidis,
who found an advantageous formulation of the evolution equations and
K.~Glampedakis for useful comments. This work
was supported by the Deutsche Forschungsgemeinschaft DFG (German Research
Foundation) via SFB/TR7, EG was funded by EGO via the VESF fellowship program.

\bibliographystyle{apsrev}
\bibliography{references}

\begin{thebibliography}{55}
\expandafter\ifx\csname natexlab\endcsname\relax\def\natexlab#1{#1}\fi
\expandafter\ifx\csname bibnamefont\endcsname\relax
  \def\bibnamefont#1{#1}\fi
\expandafter\ifx\csname bibfnamefont\endcsname\relax
  \def\bibfnamefont#1{#1}\fi
\expandafter\ifx\csname citenamefont\endcsname\relax
  \def\citenamefont#1{#1}\fi
\expandafter\ifx\csname url\endcsname\relax
  \def\url#1{\texttt{#1}}\fi
\expandafter\ifx\csname urlprefix\endcsname\relax\def\urlprefix{URL }\fi
\providecommand{\bibinfo}[2]{#2}
\providecommand{\eprint}[2][]{\url{#2}}

\bibitem[{\citenamefont{{Chandrasekhar}}(1970)}]{1970ApJ...161..561C}
\bibinfo{author}{\bibfnamefont{S.}~\bibnamefont{{Chandrasekhar}}},
  \bibinfo{journal}{\apj} \textbf{\bibinfo{volume}{161}}, \bibinfo{pages}{561}
  (\bibinfo{year}{1970}).

\bibitem[{\citenamefont{{Friedman} and
  {Schutz}}(1978{\natexlab{a}})}]{1978ApJ...221..937F}
\bibinfo{author}{\bibfnamefont{J.~L.} \bibnamefont{{Friedman}}}
  \bibnamefont{and} \bibinfo{author}{\bibfnamefont{B.~F.}
  \bibnamefont{{Schutz}}}, \bibinfo{journal}{\apj}
  \textbf{\bibinfo{volume}{221}}, \bibinfo{pages}{937}
  (\bibinfo{year}{1978}{\natexlab{a}}).

\bibitem[{\citenamefont{{Friedman} and
  {Schutz}}(1978{\natexlab{b}})}]{1978ApJ...222..281F}
\bibinfo{author}{\bibfnamefont{J.~L.} \bibnamefont{{Friedman}}}
  \bibnamefont{and} \bibinfo{author}{\bibfnamefont{B.~F.}
  \bibnamefont{{Schutz}}}, \bibinfo{journal}{\apj}
  \textbf{\bibinfo{volume}{222}}, \bibinfo{pages}{281}
  (\bibinfo{year}{1978}{\natexlab{b}}).

\bibitem[{\citenamefont{{Dimmelmeier} et~al.}(2002)\citenamefont{{Dimmelmeier},
  {Font}, and {M{\"u}ller}}}]{2002A&A...393..523D}
\bibinfo{author}{\bibfnamefont{H.}~\bibnamefont{{Dimmelmeier}}},
  \bibinfo{author}{\bibfnamefont{J.~A.} \bibnamefont{{Font}}},
  \bibnamefont{and}
  \bibinfo{author}{\bibfnamefont{E.}~\bibnamefont{{M{\"u}ller}}},
  \bibinfo{journal}{\aap} \textbf{\bibinfo{volume}{393}}, \bibinfo{pages}{523}
  (\bibinfo{year}{2002}).

\bibitem[{\citenamefont{{Ott} et~al.}(2004)\citenamefont{{Ott}, {Burrows},
  {Livne}, and {Walder}}}]{2004ApJ...600..834O}
\bibinfo{author}{\bibfnamefont{C.~D.} \bibnamefont{{Ott}}},
  \bibinfo{author}{\bibfnamefont{A.}~\bibnamefont{{Burrows}}},
  \bibinfo{author}{\bibfnamefont{E.}~\bibnamefont{{Livne}}}, \bibnamefont{and}
  \bibinfo{author}{\bibfnamefont{R.}~\bibnamefont{{Walder}}},
  \bibinfo{journal}{\apj} \textbf{\bibinfo{volume}{600}}, \bibinfo{pages}{834}
  (\bibinfo{year}{2004}).

\bibitem[{\citenamefont{{Rasio} and {Shapiro}}(1994)}]{1994ApJ...432..242R}
\bibinfo{author}{\bibfnamefont{F.~A.} \bibnamefont{{Rasio}}} \bibnamefont{and}
  \bibinfo{author}{\bibfnamefont{S.~L.} \bibnamefont{{Shapiro}}},
  \bibinfo{journal}{\apj} \textbf{\bibinfo{volume}{432}}, \bibinfo{pages}{242}
  (\bibinfo{year}{1994}).

\bibitem[{\citenamefont{Shibata and Uryu}(2000)}]{Shibata:1999wm}
\bibinfo{author}{\bibfnamefont{M.}~\bibnamefont{Shibata}} \bibnamefont{and}
  \bibinfo{author}{\bibfnamefont{K.}~\bibnamefont{Uryu}},
  \bibinfo{journal}{\prd} \textbf{\bibinfo{volume}{61}},
  \bibinfo{pages}{064001} (\bibinfo{year}{2000}).

\bibitem[{\citenamefont{{Rasio} and {Shapiro}}(1999)}]{Rasio:1999zr}
\bibinfo{author}{\bibfnamefont{F.~A.} \bibnamefont{{Rasio}}} \bibnamefont{and}
  \bibinfo{author}{\bibfnamefont{S.~L.} \bibnamefont{{Shapiro}}},
  \bibinfo{journal}{\cqg} \textbf{\bibinfo{volume}{16}}, \bibinfo{pages}{1}
  (\bibinfo{year}{1999}).

\bibitem[{\citenamefont{{Rezzolla} et~al.}(2000)\citenamefont{{Rezzolla},
  {Lamb}, and {Shapiro}}}]{2000ApJ...531L.139R}
\bibinfo{author}{\bibfnamefont{L.}~\bibnamefont{{Rezzolla}}},
  \bibinfo{author}{\bibfnamefont{F.~K.} \bibnamefont{{Lamb}}},
  \bibnamefont{and} \bibinfo{author}{\bibfnamefont{S.~L.}
  \bibnamefont{{Shapiro}}}, \bibinfo{journal}{\apjl}
  \textbf{\bibinfo{volume}{531}}, \bibinfo{pages}{L139} (\bibinfo{year}{2000}).

\bibitem[{\citenamefont{{Levin} and {Ushomirsky}}(2001)}]{2001MNRAS.324..917L}
\bibinfo{author}{\bibfnamefont{Y.}~\bibnamefont{{Levin}}} \bibnamefont{and}
  \bibinfo{author}{\bibfnamefont{G.}~\bibnamefont{{Ushomirsky}}},
  \bibinfo{journal}{\mnras} \textbf{\bibinfo{volume}{324}},
  \bibinfo{pages}{917} (\bibinfo{year}{2001}).

\bibitem[{\citenamefont{{Stergioulas} and {Font}}(2001)}]{2001PhRvL..86.1148S}
\bibinfo{author}{\bibfnamefont{N.}~\bibnamefont{{Stergioulas}}}
  \bibnamefont{and} \bibinfo{author}{\bibfnamefont{J.~A.}
  \bibnamefont{{Font}}}, \bibinfo{journal}{\prl} \textbf{\bibinfo{volume}{86}},
  \bibinfo{pages}{1148} (\bibinfo{year}{2001}).

\bibitem[{\citenamefont{{S{\'a}}}(2004)}]{2004PhRvD..69h4001S}
\bibinfo{author}{\bibfnamefont{P.~M.} \bibnamefont{{S{\'a}}}},
  \bibinfo{journal}{\prd} \textbf{\bibinfo{volume}{69}},
  \bibinfo{pages}{084001} (\bibinfo{year}{2004}).

\bibitem[{\citenamefont{{Shapiro}}(2000)}]{2000ApJ...544..397S}
\bibinfo{author}{\bibfnamefont{S.~L.} \bibnamefont{{Shapiro}}},
  \bibinfo{journal}{\apj} \textbf{\bibinfo{volume}{544}}, \bibinfo{pages}{397}
  (\bibinfo{year}{2000}).

\bibitem[{\citenamefont{{Cook} et~al.}(2003)\citenamefont{{Cook}, {Shapiro},
  and {Stephens}}}]{2003ApJ...599.1272C}
\bibinfo{author}{\bibfnamefont{J.~N.} \bibnamefont{{Cook}}},
  \bibinfo{author}{\bibfnamefont{S.~L.} \bibnamefont{{Shapiro}}},
  \bibnamefont{and} \bibinfo{author}{\bibfnamefont{B.~C.}
  \bibnamefont{{Stephens}}}, \bibinfo{journal}{\apj}
  \textbf{\bibinfo{volume}{599}}, \bibinfo{pages}{1272} (\bibinfo{year}{2003}).

\bibitem[{\citenamefont{{Hegyi}}(1977)}]{1977ApJ...217..244H}
\bibinfo{author}{\bibfnamefont{D.~J.} \bibnamefont{{Hegyi}}},
  \bibinfo{journal}{\apj} \textbf{\bibinfo{volume}{217}}, \bibinfo{pages}{244}
  (\bibinfo{year}{1977}).

\bibitem[{\citenamefont{Liu and Shapiro}(2004)}]{liu:044009}
\bibinfo{author}{\bibfnamefont{Y.~T.} \bibnamefont{Liu}} \bibnamefont{and}
  \bibinfo{author}{\bibfnamefont{S.~L.} \bibnamefont{Shapiro}},
  \bibinfo{journal}{\prd} \textbf{\bibinfo{volume}{69}},
  \bibinfo{pages}{044009} (\bibinfo{year}{2004}).

\bibitem[{\citenamefont{{Karino} and {Eriguchi}}(2002)}]{2002ApJ...578..413K}
\bibinfo{author}{\bibfnamefont{S.}~\bibnamefont{{Karino}}} \bibnamefont{and}
  \bibinfo{author}{\bibfnamefont{Y.}~\bibnamefont{{Eriguchi}}},
  \bibinfo{journal}{\apj} \textbf{\bibinfo{volume}{578}}, \bibinfo{pages}{413}
  (\bibinfo{year}{2002}).

\bibitem[{\citenamefont{{Stavridis} et~al.}(2007)\citenamefont{{Stavridis},
  {Passamonti}, and {Kokkotas}}}]{2007PhRvD..75f4019S}
\bibinfo{author}{\bibfnamefont{A.}~\bibnamefont{{Stavridis}}},
  \bibinfo{author}{\bibfnamefont{A.}~\bibnamefont{{Passamonti}}},
  \bibnamefont{and}
  \bibinfo{author}{\bibfnamefont{K.}~\bibnamefont{{Kokkotas}}},
  \bibinfo{journal}{\prd} \textbf{\bibinfo{volume}{75}},
  \bibinfo{pages}{064019} (\bibinfo{year}{2007}).

\bibitem[{\citenamefont{{Passamonti} et~al.}(2008)\citenamefont{{Passamonti},
  {Stavridis}, and {Kokkotas}}}]{2008PhRvD..77b4029P}
\bibinfo{author}{\bibfnamefont{A.}~\bibnamefont{{Passamonti}}},
  \bibinfo{author}{\bibfnamefont{A.}~\bibnamefont{{Stavridis}}},
  \bibnamefont{and} \bibinfo{author}{\bibfnamefont{K.~D.}
  \bibnamefont{{Kokkotas}}}, \bibinfo{journal}{\prd}
  \textbf{\bibinfo{volume}{77}}, \bibinfo{pages}{024029}
  (\bibinfo{year}{2008}).

\bibitem[{\citenamefont{Stergioulas et~al.}(2004)\citenamefont{Stergioulas,
  Apostolatos, and Font}}]{Stergioulas:2003ep}
\bibinfo{author}{\bibfnamefont{N.}~\bibnamefont{Stergioulas}},
  \bibinfo{author}{\bibfnamefont{T.~A.} \bibnamefont{Apostolatos}},
  \bibnamefont{and} \bibinfo{author}{\bibfnamefont{J.~A.} \bibnamefont{Font}},
  \bibinfo{journal}{\mnras} \textbf{\bibinfo{volume}{352}},
  \bibinfo{pages}{1089} (\bibinfo{year}{2004}).

\bibitem[{\citenamefont{Dimmelmeier et~al.}(2006)\citenamefont{Dimmelmeier,
  Stergioulas, and Font}}]{Dimmelmeier:2005zk}
\bibinfo{author}{\bibfnamefont{H.}~\bibnamefont{Dimmelmeier}},
  \bibinfo{author}{\bibfnamefont{N.}~\bibnamefont{Stergioulas}},
  \bibnamefont{and} \bibinfo{author}{\bibfnamefont{J.~A.} \bibnamefont{Font}},
  \bibinfo{journal}{\mnras} \textbf{\bibinfo{volume}{368}},
  \bibinfo{pages}{1609} (\bibinfo{year}{2006}).

\bibitem[{\citenamefont{{Yoshida} et~al.}(2002)\citenamefont{{Yoshida},
  {Rezzolla}, {Karino}, and {Eriguchi}}}]{2002ApJ...568L..41Y}
\bibinfo{author}{\bibfnamefont{S.}~\bibnamefont{{Yoshida}}},
  \bibinfo{author}{\bibfnamefont{L.}~\bibnamefont{{Rezzolla}}},
  \bibinfo{author}{\bibfnamefont{S.}~\bibnamefont{{Karino}}}, \bibnamefont{and}
  \bibinfo{author}{\bibfnamefont{Y.}~\bibnamefont{{Eriguchi}}},
  \bibinfo{journal}{\apjl} \textbf{\bibinfo{volume}{568}}, \bibinfo{pages}{L41}
  (\bibinfo{year}{2002}).

\bibitem[{\citenamefont{{Shibata} et~al.}(2002)\citenamefont{{Shibata},
  {Karino}, and {Eriguchi}}}]{2002MNRAS.334L..27S}
\bibinfo{author}{\bibfnamefont{M.}~\bibnamefont{{Shibata}}},
  \bibinfo{author}{\bibfnamefont{S.}~\bibnamefont{{Karino}}}, \bibnamefont{and}
  \bibinfo{author}{\bibfnamefont{Y.}~\bibnamefont{{Eriguchi}}},
  \bibinfo{journal}{\mnras} \textbf{\bibinfo{volume}{334}},
  \bibinfo{pages}{L27} (\bibinfo{year}{2002}).

\bibitem[{\citenamefont{{Shibata} and {Sekiguchi}}(2005)}]{2005PhRvD..71b4014S}
\bibinfo{author}{\bibfnamefont{M.}~\bibnamefont{{Shibata}}} \bibnamefont{and}
  \bibinfo{author}{\bibfnamefont{Y.-I.} \bibnamefont{{Sekiguchi}}},
  \bibinfo{journal}{\prd} \textbf{\bibinfo{volume}{71}},
  \bibinfo{pages}{024014} (\bibinfo{year}{2005}).

\bibitem[{\citenamefont{{Centrella} et~al.}(2001)\citenamefont{{Centrella},
  {New}, {Lowe}, and {Brown}}}]{2001ApJ...550L.193C}
\bibinfo{author}{\bibfnamefont{J.~M.} \bibnamefont{{Centrella}}},
  \bibinfo{author}{\bibfnamefont{K.~C.~B.} \bibnamefont{{New}}},
  \bibinfo{author}{\bibfnamefont{L.~L.} \bibnamefont{{Lowe}}},
  \bibnamefont{and} \bibinfo{author}{\bibfnamefont{J.~D.}
  \bibnamefont{{Brown}}}, \bibinfo{journal}{\apjl}
  \textbf{\bibinfo{volume}{550}}, \bibinfo{pages}{L193} (\bibinfo{year}{2001}).

\bibitem[{\citenamefont{{Saijo} et~al.}(2003)\citenamefont{{Saijo},
  {Baumgarte}, and {Shapiro}}}]{2003ApJ...595..352S}
\bibinfo{author}{\bibfnamefont{M.}~\bibnamefont{{Saijo}}},
  \bibinfo{author}{\bibfnamefont{T.~W.} \bibnamefont{{Baumgarte}}},
  \bibnamefont{and} \bibinfo{author}{\bibfnamefont{S.~L.}
  \bibnamefont{{Shapiro}}}, \bibinfo{journal}{\apj}
  \textbf{\bibinfo{volume}{595}}, \bibinfo{pages}{352} (\bibinfo{year}{2003}).

\bibitem[{\citenamefont{{Ou} and {Tohline}}(2006)}]{2006ApJ...651.1068O}
\bibinfo{author}{\bibfnamefont{S.}~\bibnamefont{{Ou}}} \bibnamefont{and}
  \bibinfo{author}{\bibfnamefont{J.~E.} \bibnamefont{{Tohline}}},
  \bibinfo{journal}{\apj} \textbf{\bibinfo{volume}{651}}, \bibinfo{pages}{1068}
  (\bibinfo{year}{2006}).

\bibitem[{\citenamefont{{Watts} et~al.}(2005)\citenamefont{{Watts},
  {Andersson}, and {Jones}}}]{2005ApJ...618L..37W}
\bibinfo{author}{\bibfnamefont{A.~L.} \bibnamefont{{Watts}}},
  \bibinfo{author}{\bibfnamefont{N.}~\bibnamefont{{Andersson}}},
  \bibnamefont{and} \bibinfo{author}{\bibfnamefont{D.~I.}
  \bibnamefont{{Jones}}}, \bibinfo{journal}{\apjl}
  \textbf{\bibinfo{volume}{618}}, \bibinfo{pages}{L37} (\bibinfo{year}{2005}).

\bibitem[{\citenamefont{{Saijo} and {Yoshida}}(2006)}]{2006MNRAS.368.1429S}
\bibinfo{author}{\bibfnamefont{M.}~\bibnamefont{{Saijo}}} \bibnamefont{and}
  \bibinfo{author}{\bibfnamefont{S.}~\bibnamefont{{Yoshida}}},
  \bibinfo{journal}{\mnras} \textbf{\bibinfo{volume}{368}},
  \bibinfo{pages}{1429} (\bibinfo{year}{2006}).

\bibitem[{\citenamefont{{Komatsu}
  et~al.}(1989{\natexlab{a}})\citenamefont{{Komatsu}, {Eriguchi}, and
  {Hachisu}}}]{Komatsu1989MNRASA}
\bibinfo{author}{\bibfnamefont{H.}~\bibnamefont{{Komatsu}}},
  \bibinfo{author}{\bibfnamefont{Y.}~\bibnamefont{{Eriguchi}}},
  \bibnamefont{and}
  \bibinfo{author}{\bibfnamefont{I.}~\bibnamefont{{Hachisu}}},
  \bibinfo{journal}{\mnras} \textbf{\bibinfo{volume}{237}},
  \bibinfo{pages}{355} (\bibinfo{year}{1989}{\natexlab{a}}).

\bibitem[{\citenamefont{{Komatsu}
  et~al.}(1989{\natexlab{b}})\citenamefont{{Komatsu}, {Eriguchi}, and
  {Hachisu}}}]{Komatsu1989MNRASB}
\bibinfo{author}{\bibfnamefont{H.}~\bibnamefont{{Komatsu}}},
  \bibinfo{author}{\bibfnamefont{Y.}~\bibnamefont{{Eriguchi}}},
  \bibnamefont{and}
  \bibinfo{author}{\bibfnamefont{I.}~\bibnamefont{{Hachisu}}},
  \bibinfo{journal}{\mnras} \textbf{\bibinfo{volume}{239}},
  \bibinfo{pages}{153} (\bibinfo{year}{1989}{\natexlab{b}}).

\bibitem[{\citenamefont{{Kokkotas} and
  {Vavoulidis}}(2005)}]{2005JPhCS...8...71K}
\bibinfo{author}{\bibfnamefont{K.~D.} \bibnamefont{{Kokkotas}}}
  \bibnamefont{and}
  \bibinfo{author}{\bibfnamefont{M.}~\bibnamefont{{Vavoulidis}}},
  \bibinfo{journal}{\jpcs} \textbf{\bibinfo{volume}{8}}, \bibinfo{pages}{71}
  (\bibinfo{year}{2005}).

\bibitem[{\citenamefont{{Vavoulidis}}(2007)}]{Vavoul2007}
\bibinfo{author}{\bibfnamefont{M.}~\bibnamefont{{Vavoulidis}}},
  \bibinfo{journal}{Ph.D.~Thesis, Aristotle University of Thessaloniki}
  (\bibinfo{year}{2007}).

\bibitem[{\citenamefont{{Lockitch} and {Friedman}}(1999)}]{1999ApJ...521..764L}
\bibinfo{author}{\bibfnamefont{K.~H.} \bibnamefont{{Lockitch}}}
  \bibnamefont{and} \bibinfo{author}{\bibfnamefont{J.~L.}
  \bibnamefont{{Friedman}}}, \bibinfo{journal}{\apj}
  \textbf{\bibinfo{volume}{521}}, \bibinfo{pages}{764} (\bibinfo{year}{1999}).

\bibitem[{\citenamefont{{Stavridis} and
  {Kokkotas}}(2005)}]{2005IJMPD..14..543S}
\bibinfo{author}{\bibfnamefont{A.}~\bibnamefont{{Stavridis}}} \bibnamefont{and}
  \bibinfo{author}{\bibfnamefont{K.~D.} \bibnamefont{{Kokkotas}}},
  \bibinfo{journal}{\ijmp} \textbf{\bibinfo{volume}{14}}, \bibinfo{pages}{543}
  (\bibinfo{year}{2005}).

\bibitem[{\citenamefont{{Stergioulas}}(1995)}]{rns-v1.1}
\bibinfo{author}{\bibfnamefont{N.}~\bibnamefont{{Stergioulas}}}
  (\bibinfo{year}{1995}), \urlprefix\url{http://www.gravity.phys.uwm.edu/rns/}.

\bibitem[{\citenamefont{{Stergioulas} and
  {Friedman}}(1995)}]{1995ApJ...444..306S}
\bibinfo{author}{\bibfnamefont{N.}~\bibnamefont{{Stergioulas}}}
  \bibnamefont{and} \bibinfo{author}{\bibfnamefont{J.~L.}
  \bibnamefont{{Friedman}}}, \bibinfo{journal}{\apj}
  \textbf{\bibinfo{volume}{444}}, \bibinfo{pages}{306} (\bibinfo{year}{1995}).

\bibitem[{\citenamefont{Gaertig and Kokkotas}(2008)}]{PhysRevD.78.064063}
\bibinfo{author}{\bibfnamefont{E.}~\bibnamefont{Gaertig}} \bibnamefont{and}
  \bibinfo{author}{\bibfnamefont{K.~D.} \bibnamefont{Kokkotas}},
  \bibinfo{journal}{\prd} \textbf{\bibinfo{volume}{78}},
  \bibinfo{pages}{064063} (\bibinfo{year}{2008}).

\bibitem[{\citenamefont{{Gaertig} and {Kokkotas}}(2009)}]{Gaertig:2009rm}
\bibinfo{author}{\bibfnamefont{E.}~\bibnamefont{{Gaertig}}} \bibnamefont{and}
  \bibinfo{author}{\bibfnamefont{K.~D.} \bibnamefont{{Kokkotas}}},
  \bibinfo{journal}{\prd} \textbf{\bibinfo{volume}{80}},
  \bibinfo{pages}{064026} (\bibinfo{year}{2009}).

\bibitem[{\citenamefont{{Ozel}}(2006)}]{Ozel:2006fk}
\bibinfo{author}{\bibfnamefont{F.}~\bibnamefont{{Ozel}}},
  \bibinfo{journal}{\nat} \textbf{\bibinfo{volume}{441}}, \bibinfo{pages}{1115}
  (\bibinfo{year}{2006}).

\bibitem[{\citenamefont{{Ozel} et~al.}(2009)\citenamefont{{Ozel}, {Guver}, and
  {Psaltis}}}]{Ozel:2008lr}
\bibinfo{author}{\bibfnamefont{F.}~\bibnamefont{{Ozel}}},
  \bibinfo{author}{\bibfnamefont{T.}~\bibnamefont{{Guver}}}, \bibnamefont{and}
  \bibinfo{author}{\bibfnamefont{D.}~\bibnamefont{{Psaltis}}},
  \bibinfo{journal}{\apj} \textbf{\bibinfo{volume}{693}}, \bibinfo{pages}{1775}
  (\bibinfo{year}{2009}).

\bibitem[{\citenamefont{Sotani and Kokkotas}(2004)}]{PhysRevD.70.084026}
\bibinfo{author}{\bibfnamefont{H.}~\bibnamefont{Sotani}} \bibnamefont{and}
  \bibinfo{author}{\bibfnamefont{K.~D.} \bibnamefont{Kokkotas}},
  \bibinfo{journal}{\prd} \textbf{\bibinfo{volume}{70}},
  \bibinfo{pages}{084026} (\bibinfo{year}{2004}).

\bibitem[{\citenamefont{{Diaz Alonso} and {Ibanez
  Cabanell}}(1985)}]{1985ApJ...291..308D}
\bibinfo{author}{\bibfnamefont{J.}~\bibnamefont{{Diaz Alonso}}}
  \bibnamefont{and} \bibinfo{author}{\bibfnamefont{J.~M.} \bibnamefont{{Ibanez
  Cabanell}}}, \bibinfo{journal}{\apj} \textbf{\bibinfo{volume}{291}},
  \bibinfo{pages}{308} (\bibinfo{year}{1985}).

\bibitem[{\citenamefont{{Arnett} and {Bowers}}(1977)}]{Arnett:1977yq}
\bibinfo{author}{\bibfnamefont{W.~D.} \bibnamefont{{Arnett}}} \bibnamefont{and}
  \bibinfo{author}{\bibfnamefont{R.~L.} \bibnamefont{{Bowers}}},
  \bibinfo{journal}{\apjs} \textbf{\bibinfo{volume}{33}}, \bibinfo{pages}{415}
  (\bibinfo{year}{1977}).

\bibitem[{\citenamefont{Bertil~Gustafsson and
  Oliger}(1996)}]{KreissOliger199602}
\bibinfo{author}{\bibfnamefont{H.-O.~K.} \bibnamefont{Bertil~Gustafsson}}
  \bibnamefont{and} \bibinfo{author}{\bibfnamefont{J.}~\bibnamefont{Oliger}},
  \emph{\bibinfo{title}{Time Dependent Problems and Difference Methods}}
  (\bibinfo{publisher}{John Wiley {\&} Sons}, \bibinfo{year}{1996}).

\bibitem[{\citenamefont{{Kastaun}}(2008)}]{2008PhRvD..77l4019K}
\bibinfo{author}{\bibfnamefont{W.}~\bibnamefont{{Kastaun}}},
  \bibinfo{journal}{\prd} \textbf{\bibinfo{volume}{77}},
  \bibinfo{pages}{124019} (\bibinfo{year}{2008}).

\bibitem[{\citenamefont{{Kr\"uger}}(2009)}]{Kruger:2009dq}
\bibinfo{author}{\bibfnamefont{C.}~\bibnamefont{{Kr\"uger}}},
  \bibinfo{journal}{Diploma Thesis, Eberhard-Karls University of T\"ubingen}
  (\bibinfo{year}{2009}).

\bibitem[{\citenamefont{{Zink}}()}]{Zink:yq}
\bibinfo{author}{\bibfnamefont{B.}~\bibnamefont{{Zink}}},
  \bibinfo{note}{private communication}.

\bibitem[{\citenamefont{{Baumgarte} et~al.}(2000)\citenamefont{{Baumgarte},
  {Shapiro}, and {Shibata}}}]{Baumgarte:2000rr}
\bibinfo{author}{\bibfnamefont{T.~W.} \bibnamefont{{Baumgarte}}},
  \bibinfo{author}{\bibfnamefont{S.~L.} \bibnamefont{{Shapiro}}},
  \bibnamefont{and}
  \bibinfo{author}{\bibfnamefont{M.}~\bibnamefont{{Shibata}}},
  \bibinfo{journal}{\apjl} \textbf{\bibinfo{volume}{528}}, \bibinfo{pages}{L29}
  (\bibinfo{year}{2000}).

\bibitem[{\citenamefont{{Andersson}}(1998)}]{1998ApJ...502..708A}
\bibinfo{author}{\bibfnamefont{N.}~\bibnamefont{{Andersson}}},
  \bibinfo{journal}{\apj} \textbf{\bibinfo{volume}{502}}, \bibinfo{pages}{708}
  (\bibinfo{year}{1998}).

\bibitem[{\citenamefont{Karino et~al.}(2001)\citenamefont{Karino, Yoshida, and
  Eriguchi}}]{PhysRevD.64.024003}
\bibinfo{author}{\bibfnamefont{S.}~\bibnamefont{Karino}},
  \bibinfo{author}{\bibfnamefont{S.}~\bibnamefont{Yoshida}}, \bibnamefont{and}
  \bibinfo{author}{\bibfnamefont{Y.}~\bibnamefont{Eriguchi}},
  \bibinfo{journal}{\prd} \textbf{\bibinfo{volume}{64}},
  \bibinfo{pages}{024003} (\bibinfo{year}{2001}).

\bibitem[{\citenamefont{{Yoshida} et~al.}(2005)\citenamefont{{Yoshida},
  {Yoshida}, and {Eriguchi}}}]{2005MNRAS.356..217Y}
\bibinfo{author}{\bibfnamefont{S.}~\bibnamefont{{Yoshida}}},
  \bibinfo{author}{\bibfnamefont{S.}~\bibnamefont{{Yoshida}}},
  \bibnamefont{and}
  \bibinfo{author}{\bibfnamefont{Y.}~\bibnamefont{{Eriguchi}}},
  \bibinfo{journal}{\mnras} \textbf{\bibinfo{volume}{356}},
  \bibinfo{pages}{217} (\bibinfo{year}{2005}).

\bibitem[{\citenamefont{{Kojima}}(1998)}]{1998MNRAS.293...49K}
\bibinfo{author}{\bibfnamefont{Y.}~\bibnamefont{{Kojima}}},
  \bibinfo{journal}{\mnras} \textbf{\bibinfo{volume}{293}}, \bibinfo{pages}{49}
  (\bibinfo{year}{1998}).

\bibitem[{\citenamefont{{Beyer} and {Kokkotas}}(1999)}]{1999MNRAS.308..745B}
\bibinfo{author}{\bibfnamefont{H.~R.} \bibnamefont{{Beyer}}} \bibnamefont{and}
  \bibinfo{author}{\bibfnamefont{K.~D.} \bibnamefont{{Kokkotas}}},
  \bibinfo{journal}{\mnras} \textbf{\bibinfo{volume}{308}},
  \bibinfo{pages}{745} (\bibinfo{year}{1999}).

\bibitem[{\citenamefont{{Ruoff} and {Kokkotas}}(2001)}]{2001MNRAS.328..678R}
\bibinfo{author}{\bibfnamefont{J.}~\bibnamefont{{Ruoff}}} \bibnamefont{and}
  \bibinfo{author}{\bibfnamefont{K.~D.} \bibnamefont{{Kokkotas}}},
  \bibinfo{journal}{\mnras} \textbf{\bibinfo{volume}{328}},
  \bibinfo{pages}{678} (\bibinfo{year}{2001}).

\end{thebibliography}

\end{document}